\documentstyle{article}
\newcommand{\beq}{\begin{equation}}
\newcommand{\eeq}{\end{equation}}
\newcommand{\beqa}{\begin{eqnarray}}
\newcommand{\eeqa}{\end{eqnarray}}
\newcommand{\half}{\frac{1}{2}}
\newcommand{\x}{{\bf x}}
\newcommand{\r}{{\bf r}}
\newcommand{\z}{{\bf z}}
\renewcommand{\a}{{\bf a}}
\newcommand{\w}{{\bf w}}
\renewcommand{\xi}{\x_i}
\newcommand{\ri}{\r_i}
\newcommand{\zi}{\z_i}
\newcommand{\wi}{\w_i}
\newcommand{\ai}{\a_i}

\newcommand{\rj}{\r_j}
\newcommand{\zj}{\z_j}
\newcommand{\wj}{{\bf w}_j}
\newcommand{\aj}{\a_j}
\newcommand{\xij}{\x_{ij}}
\newcommand{\Gij}{G_{ij}}

\newcommand{\Ginv}{G^{-1}}
\newcommand{\Ginvij}{\Ginv_{\;\, ij}}
\newcommand{\sig}{\sigma}
\newcommand{\rsig}{\r_{\sig}}
\newcommand{\zsig}{\z_{\sig}}
\newcommand{\asig}{\a_{\sig}}
\newcommand{\sumsig}{\sum_{\sig}}
\newcommand{\sumsigi}{\sum_{\sig\ni i}}

\newcommand{\sumsigij}{\sum_{\sig\ni i,j}}
\newcommand{\sumsigijk}{\sum_{\sig\ni i,j,k}}
\newcommand{\sumisig}{\sum_{i\in\sig}}
\newcommand{\sumjsig}{\sum_{j\in\sig}}

\newcommand{\erf}{\mbox{erf}}
\newcommand{\erfc}{\mbox{erfc}}
\newcommand{\sqtopi}{\sqrt{\frac{2}{\pi}}}
\newcommand{\J}{{\bf J}}
\input{psfig}

\begin{document}
\begin{titlepage}
\begin{flushright}
December 1993\\ LU TP 93--15\\Revised version
\end{flushright}
\vspace{0.05in}
\LARGE
\begin{center}
{\bf A Variational Approach to the Structure and Thermodynamics of Linear
Polyelectrolytes with Coulomb and Screened Coulomb Interactions}\\
\vspace{.15in}
\large
Bo J\"{o}nsson\footnote{fk2boj@grosz.fkem2.lth.se}\\
\vspace{0.05in}
Physical Chemistry 2, Chemical Center, University of Lund\\ Box 124, S-221 00 Lund, Sweden\\
\vspace{0.15in}
Carsten Peterson\footnote{carsten@thep.lu.se} and Bo
S\"{o}derberg\footnote{bs@thep.lu.se}\\
\vspace{0.05in}
Department of Theoretical Physics, University of Lund\\ S\"{o}lvegatan 14A,
S-22362 Lund, Sweden\\
\vspace{0.15in}

{\it Journal of Physical Chemistry} {99}, 1251 (1995)

\end{center}
\normalsize

Abstract:

A variational approach, based on a discrete representation of the
chain, is used to calculate free energy and conformational properties
in polyelectrolytes. The true bond and Coulomb potentials are
approximated by a trial isotropic harmonic energy containing
force constants between
{\em all}
monomer-pairs as variational parameters. By a
judicious choice of representation and the use of incremental matrix
inversion, an efficient and fast-convergent iterative algorithm is
constructed, that optimizes the free energy.  The computational demand
scales as $N^3$ rather than $N^4$ as expected in a more naive
approach.  The method has the additional advantage that in contrast to
Monte Carlo calculations the entropy is easily computed. An analysis
of the high and low temperature limits is given. Also, the variational
formulation is shown to respect the appropriate virial identities.
The accuracy of the approximations introduced are tested against Monte
Carlo simulations for problem sizes ranging from $N=20$ to 1024. Very
good accuracy is obtained for chains with unscreened Coulomb
interactions.
The addition of salt is described through a screened Coulomb
interaction, for which the accuracy in a certain parameter range turns
out to be inferior to the unscreened case. The reason is that the
harmonic variational
Ansatz becomes less efficient with shorter range
interactions.

As a by-product a very efficient Monte Carlo algorithm was developed
for comparisons, providing high statistics data for very large sizes
-- 2048 monomers. The Monte Carlo results are also used to examine
scaling properties, based on low-$T$ approximations to end-end and
monomer-monomer separations. It is argued that the former increases
faster than linearly with the number of bonds.

\end{titlepage}

\normalsize

\section{Introduction}

Polymers and polymer solutions play a profound role in our daily life,
biologically and technologically. This is of course one reason for the
intense theoretical studies of polymers, but they have also because of
their very nature been a challenge to theoreticians.  Much theoretical
work has been done in order to obtain a general understanding of
neutral polymers, either in melts or in solution \cite{gen,cloiz}.
Polyelectrolytes, on the other hand, have been less thoroughly
investigated, despite their importance in many technical applications
-- for example in glue production or in food industry, for promoting
flocculation or in pulp drying \cite{jones,schee,gut}. The term
polyelectrolyte is sometimes used as a collective name for any highly
charged aggregate. However, here we will restrict the meaning to a
flexible molecule with several or many charged or chargeable sites;
poly-L-glutamic acid, polyamines and polysaccharides are some typical
representatives.

The conformation of a flexible polyelectrolyte is a result of the competition
between the covalent bonding forces, electrostatic interactions as well as more
specific short ranged interactions. For example, poly-L-glutamic acid and
several polysaccharides undergo helix to coil transition as a function of pH
\cite{snipp,ol}. This transition is obviously governed by electrostatic 
forces -- 
similar structural transitions are seen for DNA \cite{dna}. Undoubtedly, both
the polymer nature and the interaction between charged amino acids play an
important role for the folding of a protein, as well as other solvent averaged
forces. The present study should be seen as an attempt to approach the folding
problem using what turns out to be a very powerful statistical mechanical
variational technique. In the first step we will limit the study to linear
polyelectrolytes in salt solution.

Variational methods are standard techniques in quantum mechanics, but less so in
statistical mechanics, although variational principles were formulated many
years ago \cite{feyn}. One type of variational formulations starts off from an
approximate free energy, which is optimized with respect to the particle density
\cite{evans}. For polymers a more fundamental approach is possible, by
introducing variational
parameters directly into the appropriate Hamiltonian. In the past this 
route has been followed in a number of polymer studies 
\cite{cloiz2,edw,qian} and it has recently been revitalized by several 
groups \cite{brat,pod}. To the best of our knowledge, all 
these calculations have been concerned with continous chains and
only one or at most a few variational parameters have been optimized.

The present approach, of which some results were already published
\cite{boj}, is inspired by refs. \cite{cloiz1,mez1,mez2}.  It uses a
discrete representation of the polymer, which not only allows us to
investigate linear or cyclic polymers, but also polymers of arbitrary
topology.  Thus, e.g., a hyperbranched dendrite structure can easily
be handled within this formalism.
It relies on a variational Ansatz in the form of a generic Gaussian distribution,
with adjustable force constants between {\em every} pair of monomers
(not to be confused with a Brownian model with forces only between nearest neighbours).
Thus, the number of variational parameters
is proportional to $N^2$, where $N$ is the number of interacting
units.  In general, the variational approach is expected to be most
accurate at high dimensions \cite{cloiz1,mez1}.  Apparantly, this has
discouraged the community from pushing the approach for
three-dimensional polymers into a numerical confrontation. Also, using
the method in a naive way would give a
computational effort
scaling like $N^4$, which would make the method less tractable for large sizes. We have
also found empirically that such a naive implementation is plagued
with bad convergence properties. In previous work \cite{boj} an
algorithm was developed that lowers the computational costs to $N^3$
with controlled and nice convergence properties.

The high and low $T$ limits of the variational approach, which are accessible
with analytical means, are derived in this paper. The low $T$ expansions
for various quantities yield results that very well approximate what
emerges from the corresponding expansions in the exact theory. The fact that
the temperature of interest (room temperature) is fairly low on the
temperature scale partly explains the success of initial numerical
explorations of the variational approach \cite{boj} and motivates further
studies.

Besides being more realistic, the discrete chain also offers the possibility of
direct comparison with Monte Carlo (MC) simulations, thereby giving an
important indication of the accuracy.  The final output of the present
variational calculation is of course the minimized free energy, but one also
obtains a matrix with all possible monomer-monomer correlations within the
chain. This matrix is the starting point for the calculation of end-end and
monomer-monomer separations and different kinds of angular correlations.

MC simulations of flexible polyelectrolytes have only recently appeared in the
literature and then limited to rather short chains
\cite{baum,brend,carn,vall,higgs,gran,reed,sass,stev}, the longest chains
being of the order of a few hundred monomers.  So far simulations have
dealt with both explicit representation of salt particles as well as
implicit in the form of a screened Coulomb potential. The inclusion of
short ranged interactions has also been studied as have the
conformational changes associated with the titration of a flexible
polyelectrolyte \cite{sass}. Most simulations have been carried out in
the canonical ensemble, although the latter problem required the use
of grand canonical MC simulations \cite{reed,sass}. The accuracy of
the screened Coulomb potential has been investigated by several
people and found to be an excellent approximation for not too high
polyelectrolyte charge densities and in the absence of any multivalent
ions \cite{vall,wood}. In order to obtain high statistics MC results a
pivot algorithm and a recently developed hybrid scheme \cite{irb} were
used.  As a by-product these simulation results were also used to
examine certain scaling properties, theoretically derived from an
approximate low $T$ expression.
In contrast to
the case of rigid bonds we find that the end-end separations scale
faster than linearly with N.

When confronting the variational approach with MC data the following
results emerge in this work:
\begin{itemize}
\item The variational approach has the unique property that it directly
yields the free energy $F$. This is in contrast to MC simulations, where F is
only indirectly accessible through elaborate integrations.  For N=20,
where comparisons are inexpensive, the variational free energy nicely agrees 
with MC results.
\item For unscreened Coulomb chains the success reported in ref. \cite{boj}
survives to even larger systems (N=1024) for configurational
properties like end-end correlations. Also for angular correlations and scaling
behaviour the variational approach reproduces MC data very well.
\item When including salt through Debye screened Coulomb potentials
the performance of the method deteriorates somewhat on the quantitative
level. Even if the qualitative configurational picture agrees with that from MC
data, the actual numbers for e.g. end-end correlations could differ up to 50
\%. We attribute this discrepancy partly to the inability of
harmonic forces to
reproduce short range interactions.
\end{itemize}

One should also mention that other approximation schemes than the
variational ones have been attempted in order to study polyelectrolyte
conformations. In the mean field approximation it is only with the
assumption of spherical symmetry that the mean field equations become
tractable, but the results are not particularly encouraging
\cite{wood}. In a cylindrical geometry, however, the mean field
approximation behaves much more satisfactorily, but at the expense of
large numerical efforts.

This paper is organized as follows: In sect. 2 the basic Coulomb
models (screened and unscreened) are presented. The variational
approach is presented in sect. 3. The high and low $T$ limits of the
variational scheme are computed with analytical methods in sect. 4. A
description of the MC method used in order to establish the quality of
the variational method is found in Sect. 5. The results from
confronting the variational approach with MC data are presented in
sect. 6. Finally in sect. 7 a brief summary and outlook is given. Most
of the detailed derivations are found in appendices -- generics about
the variational method (A), variational energies for a polyelectrolyte
(B), virial identities (C), high and low $T$ expansions (D) and zero
temperature scaling properties (E). The disposition of the material
into bulk text and appendices is such that the approach and results
can be fully understood without reading the appendices.

\section{The Model}

\subsection{The Unscreened Coulomb Chain}

In this model we consider a polyelectrolyte at infinite dilution and
without any added salt. The polyelectrolyte counterions are thus
neglected and the only electrostatic interactions are between the
charged monomers.  More explicitly, the polymer chain consists of $N$
point charges connected by harmonic oscillator (``Gaussian'')
bonds. The potential energy for a chain then takes the form
\beq
\label{E}
	\tilde{E} = \tilde{E}_G + \tilde{E}_C = \frac{k}{2} \sum_{i=1}^{N-1}
	|\tilde{\x}_{i,i+1}|^2 + \frac{q^2}{4\pi \epsilon _r \epsilon
	_0}\sum_{i<j}\frac{1}{|\tilde{\x}_{ij}|}
\eeq
Here, $\tilde{\x}_i$ the position of the $i$th charge, and
\beq
\label{rij}
	\tilde{\x}_{ij} = \tilde{\x}_i - \tilde{\x}_j
\eeq
while $q$ is the monomer charge and $\epsilon _r \epsilon _0$ is the dielectric
permittivity of the medium.  We use the tilde notation $\tilde{E}$,
$\tilde{\x}_i$, etc. for physical quantities in conventional units, and reserve
$E$, $\xi$, etc. for dimensionless ones, which will be used in the theoretical
formalism below.

The force constant can be reexpressed in terms of the $N=2$ equilibrium distance
$r_0$, given by
\beq
\label{k}
	k = \frac{q^2}{4\pi \epsilon _r \epsilon _0} \frac{1}{r_0^3}
\eeq
In terms of the dimensionless coordinates $\xi$, defined by
\beq
\label{xi}
	\tilde{\x}_i = r_0 \xi
\eeq
the energy takes the form
\beq
\label{E1}
	\tilde{E} = kr_o^2 \left ( \frac{1}{2} \sum_i |\x_{i,i+1}|^2 +
	\sum_{i<j}\frac{1}{|\xij|} \right )
\eeq
We will consider the system at a finite temperature $\tilde{T}$,
which can be similarly rescaled,
\beq
\label{T}
	T = \frac{k_B \tilde{T}}{k r_0 ^2}
\eeq
with $k_B$ the Boltzmann constant.  One then obtains for the Boltzmann
exponent the simple expression
\beq
\label{B}
	\frac{\tilde{E}}{k_B\tilde{T}} = \frac{E}{T} = \frac{1}{T}
	\left ( \half \sum_i |\x_{i,i+1}|^2 +
	\sum_{i<j}\frac{1}{|\xij|} \right )
\eeq
In other words, a polyelectrolyte at infinite dilution represents a
two-parameter model where $T$ and the number of monomers $N$ are the only two
non-trivial parameters of the system.

Unless otherwise stated the following parameter values will be used throughout
the paper; $\tilde{T}$=298K, $\epsilon_r$=78.3 and $r_0$=6{\AA}.  Obviously,
$E$ depends only on the relative positions; the global center-of-mass position
variable will have to be excluded from integrations over the coordinate 
space.

The Gaussian and Coulomb energies are subject to a {\em virial identity} (see
Appendix C), which in dimensionless units reads
\beq
\label{vir}
	2 \langle E_G \rangle - \langle E_C \rangle =3(N-1)T
\eeq
Eq. (\ref{vir}) is a useful relation for checking the correctness and
convergence behaviour in MC simulations and we find that it is in general obeyed
to 0.3\% or better.

\subsection{The Screened Coulomb Chain}

To treat a single polyelectrolyte in a solution at finite salt concentration
becomes very costly in a MC simulation, since for reasonable salt concentrations
the number of salt ions easily becomes prohibitively large, much larger than the
number of polyelectrolyte monomers.  The usual way to avoid this problem is to
preaverage the degrees of freedom of the simple salt ions for some fixed
configuration of the polyelectrolyte \cite{ber}, thus defining salt averaged
effective potentials -- this is the basis of the electrostatic contribution in
the classical DLVO potential \cite{ver}. In this way we may derive a screened
Coulomb potential from a linearisation of the Poisson-Boltzmann equation for the
salt. Eq. (\ref{E}) is then replaced by
\beq
\label{E2}
	\tilde{E} = \tilde{E}_G + \tilde{E}_C = \frac{k}{2} \sum_i
	|\tilde{\x}_{i,i+1}|^2 + \frac{q^2}{4\pi \epsilon_r
	\epsilon_0}\sum_{i<j} \frac{e^{-\tilde{\kappa}
	|\tilde{\x}_{ij}|}}{|\tilde{\x}_{ij}|}
\eeq
where $\tilde{\kappa}$ is the Debye screening length for a 1:1 salt defined as
\beq
\label{kappa}
	\tilde{\kappa} = q \sqrt{\frac{2N_A c_s}{\epsilon _r \epsilon _0 k_B T}}
\eeq

In eq. (\ref{kappa}) $c_s$ is the salt concentration in molars (M) and $N_A$ is
the Avogadro's number.  The Boltzmann factor will for the screened Coulomb
potential contain the inverse dimensionless Debye screening length, $\kappa =
r_0 \tilde{\kappa}$ as an additional parameter, and with the parameter values
given above we have for $c_s$ = 0.01 M, 0.1 M and 1.0 M the $\kappa$-values
0.1992, 0.6300 and 1.992 respectively.  Then eq. (\ref{B}) is modified into
\beq
\label{B1}
	\frac{E}{T} = \frac{1}{T} \left ( \half \sum_i
	|\x_{i,i+1}|^2 + \sum_{i<j}\frac{e^{-\kappa |\xij|}}{|\xij|}
	\right )
\eeq
The virial identity (see Appendix C) now takes the form
\beq
\label{vir1}
	2 \langle E_G \rangle - \langle E_C \rangle -\kappa \langle
	\sum_{i<j}e^{-\kappa |\xij|} \rangle =3(N-1)T
\eeq

\subsection{Relative coordinates}

In the remainder of this paper, relative coordinates will be used; instead of the
absolute monomer positions $\xi$, the {\em bond vectors} $\ri$,
\beq
	\ri \equiv \x_{i+1}-\xi , \; \; i = 1,\ldots,N-1
\eeq
will be used as the fundamental variables. In this way complications due to the
translational zero-mode are avoided; in addition the convergence of the
algorithm is considerably speeded up, especially at high temperatures.

The energy of the screened Coulomb chain will then take the following form:
\beq
	E(\r) = E_G + E_C = \half \sum_{i=1}^{N-1} \ri^2 +
	\sumsig \frac{e^{-\kappa r_{\sig}}}{r_{\sig}}
\eeq
where $\sig$ runs over contiguous non-nil sub-chains, with
\beq
	\rsig \equiv \sumisig \ri
\eeq
corresponding to the distance vector between the endpoints of the subchain.  The
unscreened chain results for $\kappa=0$.

\section{The Variational Approach}

\subsection{The Gaussian Ansatz}

In refs. \cite{mez1,mez2,boj} the variational method of refs. \cite{feyn,cloiz1}
(see Appendix A for a generic description) was revisited in the context of
discrete chains of polyelectrolytes. The approach is based on an effective
energy Ansatz $E_V$, given by
\beq
\label{E_V}
	E_V / T = \half \sum_{ij} \Ginvij \, (\ri - \ai) \cdot
	(\rj - \aj)
\eeq
where $\ai$ defines an average bond vector, around which Gaussian fluctuations
are given by the symmetric positive-definite correlation matrix $\Gij$, the
matrix inverse of which appears in the energy.

Using this effective energy, the exact free energy $F = -T \log Z$ of the
polymer is approximated from above \cite{feyn} by the variational one
\beq
\label{F1}
	\hat{F} = F_V + \langle E - E_V \rangle_V \geq F
\eeq
where $F_V = -T \log Z_V$, and $\langle \rangle_V$ refers to averages with
respect to the trial Boltzmann distribution $\exp(-E_V / T)$.

The parameters $\Gij$ and $\ai$ are to be determined such that the variational
free energy $\hat{F}$ is minimized; we note that the number of variational
parameters increases with $N$ like $N^2$.  The resulting effective Boltzmann
distribution is then used to approximate expectation values $\langle f\rangle$
by effective ones $\langle f\rangle_V$.  Thus, we have e.g.
\beqa
	\langle \ri \rangle_V & = & \ai \\ \langle \ri \cdot \rj \rangle_V & = &
	\ai \cdot \aj + 3 \Gij
\eeqa
%

For the polyelectrolyte systems treated in this study we will at high
temperatures find a unique variational solution, characterized by
$\ai=0$; this defines a {\em purely fluctuating} solution.  At low
temperatures we find in addition a {\em rigid} solution with aligned
$\ai \neq 0$. The latter is due to spontaneous symmetry-breaking; it
ceases to exist at high temperatures, but will at low enough
temperatures have the lower free energy. As discussed below, the rigid
solution can typically be disregarded at normal temperatures.

For potentials more singular than $1/r^2$, $\langle E \rangle_V$ will be
divergent, and the approach breaks down.  However, such potentials are not
physical and we do not consider this limitation of the approach a serious one.

A non-trivial result of the scaling properties of the effective energy is that
the virial identity, eqs. (\ref{vir},\ref{vir1}), will be respected by the
above variational approach (see Appendix C).

\subsection{Using Local Fluctuation Amplitudes}

The minimization of $\hat{F}$ with respect to $\Gij$ and $\ai$ gives rise to a
set of matrix equations to be solved iteratively.
%
%
These are considerably simplified, and the symmetry and positivity constraints
on $\Gij$ are automatic, if $\Gij$ is expressed as the product of a matrix and
its transpose:
\beq
	\Gij = \sum_{\mu = 1}^{N-1} z_{i\mu} z_{j\mu} = \zi \cdot \zj
\eeq

The interpretation of the local parameter $\zi$ is simple -- it is a fluctuation
amplitude for the $i$th bond vector $\ri$. We can write
\beq
\label{zi}
	\ri = \ai + \sum_{\mu} z_{i\mu} \J_{\mu}
\eeq
where each component of $\J_{\mu} \in {\cal R}^3$ is an independent Gaussian
noise variable of unit variance.

Similarly, we have for a subchain
\beq
\label{zsig}
	\rsig = \sumisig \ri \equiv \asig + \sum_{\mu} z_{\sig\mu} \J_{\mu},
\eeq
where $\asig = \sumisig \ai$ and $\zsig = \sumisig \zi$. Thus, the noise
amplitudes are additive.

The matrix inverse of $G$ can similarly be decomposed:
\beq
	\Ginvij = \wi \cdot \wj
\eeq
where $w_{i\mu}$ is the (transposed) matrix inverse of $z_{i\mu}$:
\beq
\label{wi}
	\zi \cdot \wj = \delta_{ij}
\eeq
Note that $\zi$, $\wi$ and $\zsig$ are vectors, not in $R^3$, but in $R^{N-1}$.

The equations for a local extremum of $\hat{F}(\a,\z)$ are obtained by
differentiation with respect to $\zi$ and $\ai$,
\beq
\label{dF0}
	\frac{\partial \hat{F}}{\partial \zi } = 0 \; , \;
	\frac{\partial \hat{F}}{\partial \ai } = 0
\eeq

\subsection{The Unscreened Coulomb Chain}

In terms of $\ai$ and $\zi$, the variational free energy for the pure
Coulomb chain becomes, ignoring trivial additive constants (see
Appendix B):
\beq
\label{F}
	\hat{F} = - 3T \log \det z + \half \sum_i ( 3 \zi^2 + \ai^2 )
	+ \sumsig \frac{1}{a_{\sig}} \; \erf \left ( \frac{ a_{\sig}
	}{ \sqrt{2} \; z_{\sig} } \right )
\eeq
The equations for a minimum will be
\beqa
\label{dFdz}
	\frac{\partial \hat{F}}{\partial \zi} & = & -3 T \wi + 3 \zi -
	\sqtopi \sumsigi \frac{\zsig}{z_{\sig}^3} \exp \left (
	-\frac{a_{\sig}^2}{2 z_{\sig}^2} \right ) = 0 \\
\label{dFda}
	\frac{\partial\hat{F}}{\partial\ai} & = & \ai - \sumsigi
	\frac{\asig}{a_{\sig}^3} \left [ \sqtopi
	\frac{a_{\sig}}{z_{\sig}} - \erf \left (
	\frac{a_{\sig}}{\sqrt{2}\; z_{\sig}} \right ) \right ] = 0
\eeqa
where the reciprocal vector $\wi$ is defined by eq. (\ref{wi}).

These equations allow a purely fluctuating solution with $\ai =
0$. Setting $\ai=0$ the variational free energy simplifies to
\beq
\label{FZ}
	\hat{F} = -3T \log \det z + \frac{3}{2} \sum_i \zi^2 + \sqtopi
	\sumsig \frac{1}{z_{\sig}}
\eeq
which looks very much like the energy of an $(N-1)$-dimensional
Coulomb chain with bonds $\zi$, but with an extra {\em entropy} term
(the first) preventing alignment of the ground state.
The $\z$ derivatives become
\beq
\label{namn}
	\frac{\partial \hat{F}}{\partial \zi} = -3T \wi + 3 \zi -
	\sqtopi \sumsigi \frac{\zsig}{z_{\sig}^3} = 0
\eeq

In order to bring out the structure of the variational solution, we take
the scalar product of eq. (\ref{namn}) with $\wj$ to get an expression for the force constants,
\beq
	T G^{-1}_{ij} =  \delta_{ij} - \frac{1}{3} \sqtopi \sumsigij {z_{\sig}^3}
\eeq
Thus the variational energy that minimizes the free energy has the following
structure:
\beq
E_V = \half \sum_{i=1}^{N-1} \ri^2 - \frac{1}{6} \sqtopi \sumsig \frac{r_{\sig}^2}{z_{\sig}^3}
\eeq
The first term is just the original bond term, while in the second term
the Coulomb interactions are replaced by repulsive harmonic forces, having
the right scale to give a good approximation to the Coulomb interactions
for typical distances $r_{\sig} \propto z_{\sig}$.

\subsection{The Screened Coulomb Chain}

In the case of Debye screening the expression for $\hat{F}$ is
modified to
\beqa
\label{F_s}
	\hat{F} & = & - 3T \log \det z + \half \sum_i ( 3 \zi^2 + \ai^2 ) \\
\nonumber
	& + & \sumsig \frac{1}{2 a_{\sig}} \; \exp \left ( -
	\frac{a_{\sig}^2}{2 z_{\sig}^2} \right ) \left \{
	\Psi \left ( \kappa z_{\sig} - \frac{a_{\sig}}{z_{\sig}} \right )
	-
	\Psi \left ( \kappa z_{\sig} + \frac{a_{\sig}}{z_{\sig}} \right )
	\right \}
\eeqa
where
\beq
\label{psi}
	\Psi(x) \equiv \exp(x^2/2) \; \erfc(x/\sqrt{2})
\eeq
The corresponding derivatives (cf. eq. (\ref{dFdz})) take the form
\beqa
\label{dFdz1}
	\frac{\partial \hat{F}}{\partial \zi} & = & -3 T \wi + 3 \zi
	\\
\nonumber
	& - & \sumsigi \frac{\kappa^2 \zsig}{2a_{\sig}} \exp \left ( -
	\frac{a_{\sig}^2}{2 z_{\sig}^2} \right ) \left \{
	\Psi \left ( \kappa z_{\sig} - \frac{a_{\sig}}{z_{\sig}} \right )
	-
	\Psi \left ( \kappa z_{\sig} + \frac{a_{\sig}}{z_{\sig}} \right )
	- \sqtopi \frac{a_{\sig}}{\kappa^2 z_{\sig}^3} \right
	\} = 0 \\
\label{dFda1}
	\frac{\partial \hat{F}}{\partial \ai} & = & \ai \\
\nonumber
	& - & \sumsigi \frac{\kappa \asig}{2 a_{\sig}^2} \exp \left (
	- \frac{a_{\sig}^2}{2 z_{\sig}^2} \right ) \left \{
	\Psi \left ( \kappa z_{\sig} - \frac{a_{\sig}}{z_{\sig}} \right )
	+
	\Psi \left ( \kappa z_{\sig} + \frac{a_{\sig}}{z_{\sig}} \right )
	- \sqtopi \frac{2}{\kappa z_{\sig}} \right \} = 0
\eeqa

Also here, a purely fluctuating solution is allowed. Setting $\ai=0$
the variational free energy reduces to
\beq
\label{F_sZ}
	\hat{F} = - 3T \log \det z + \frac{3}{2} \sum_i \zi^2 +
	\sumsig \left \{ \sqtopi \frac{1}{z_{\sig}} - \kappa \Psi
	\left ( \kappa z_{\sig} \right ) \right \}
\eeq
The $\z$ derivatives will be
\beq
	\frac{\partial \hat{F}}{\partial \zi} = -3 T \wi + 3 \zi -
	\sumsigi \frac{\zsig}{z_{\sig}^3} \left \{
	\sqtopi (1 - \kappa^2 z_{\sig}^2)
	+ \kappa^3 z_{\sig}^3 \Psi \left ( \kappa z_{\sig} \right )
	\right \} = 0
\eeq

\subsection{Implementation}

Due to the use of relative coordinates and of local noise amplitudes, a simple
gradient descent method with a large step-size $\epsilon$ can be used, that
gives fast convergence to a solution of eqs. (\ref{dF0})
\beq
\label{gdz1}
	\Delta \zi = - \epsilon_z \frac{\partial \hat{F}} {\partial \zi}, \;
	\Delta \ai = - \epsilon_a \frac{\partial \hat{F}}{\partial \ai},
\eeq
Further speed is gained by updating the reciprocal variables $\wi$ using
incremental matrix inversion \cite{boj} -- the increment in $\wj$ due to
$\Delta\zi$ is given (exactly) by
\beq
\label{gdom}
	\Delta \wj = - \frac{ \wi (\wj \cdot \Delta \zi)} {1 + \wi \cdot \Delta
	\zi}
\eeq
to be applied in parallel for $j$ for fixed $i$.  As a by-product the
denominator $(1+\wi\cdot\Delta\zi)$ gives the multiplicative change in
the determinant $\det z$, needed to keep track of $\hat{F}$.

In order to maintain a reasonable numerical precision, the erf-related
functions needed in the process are evaluated using carefully defined Taylor
expansions or asymptotic expansions, depending on the size of the argument.

As discussed above, the equations for a minimum are consistent with a
purely fluctuating solution $\ai = 0$.  Such a solution does indeed exist
at all $T$; furthermore it is the only solution for high enough $T$.
%
%
It turns out that for realistic choices of $T$ one is in the region where
this solution gives rise to good results (see figs. \ref{r_Tab},
\ref{r_Tcd} below). The additional solution with $\ai \neq 0$ appearing at
low $T$ is a symmetry-broken solution; to be realistic, such a solution
should show an anisotropy also in the fluctuations, i.e. different
amplitudes $\zi$ for the fluctuations parallel and transverse to the
direction defined by (the aligned) $\ai$.  This requires a more general
Ansatz than eq. (\ref{E_V}); theoretically, this will produce better low-$T$
solutions, but for the Coulomb chain it leads to equations containing
functions that are difficult to evaluate numerically, so we will not use
this possibility in this paper.  The incomplete symmetry-breaking partly 
explains the tendency for the $\a \neq 0$ solutions to produce 
inferior solutions.

For the above reasons, and for reasons of continuity, we will in the
numerical explorations use the $\a = 0$ solutions, where not otherwise
stated. This also implies faster performance since only the
$\zi$-variables, with simplified updating equations, are needed.

The complete algorithm will look as follows:
\begin{enumerate}
\item
	Initialize $\zi$ (and $\ai$ if present) randomly, suitably in the
	neighborhood of a truncated high- or low-$T$ series solution.
\item
	For each $i$:
	\begin{itemize}
	\item Update $\zi$ (and $\ai$) according
	to eqs. (\ref{gdz1}) with suitable step-sizes.
	\item Correct all $\wj$ according to eq. (\ref{gdom}).
	\end{itemize}
\item
	Check if converged; if not, go to 2.
\item
	Extract $\ai$ and $\Gij = \zi \cdot \zj$, and compute variational
	averages of interest.
\end{enumerate}
Typical step-sizes are $\epsilon_z \approx 1/6$ and $\epsilon_a \approx 1/2$.
The convergence check is done based on the rate of change and on the virial
identity.  The number of computations in each iteration step for this procedure
is proportional to $N^3$. The number of iterations required for convergence is
a slowly growing function $g(N)$.  In total, thus, the execution time of the
algorithm grows with $N$ as $N^3g(N)$. In terms of CPU requirement convergence 
of a $N=40$ chain ($\ai=0$) requires 3 seconds on a DEC Alpha workstation.

\section{High and Low T Results -- Analytical Considerations}

Before embarking on a numerical evaluation of the variational approach
with comparisons to MC results, it is interesting to see what can be
gained from studying the high and low $T$ limits, where analytical
methods can be used.

At the energy minimum, prevailing at $T = 0$, the polyelectrolyte will
form a straight line. When the temperature is increased there will be
a competition between the entropy and the repulsive Coulomb forces,
and as $T \rightarrow \infty$, the chain becomes Brownian, and the
elongated or ordered structure is gone altogether. The temperature
range of the transition from ordered to disordered structure is
$N$-dependent. As $N$ increases at fixed $T$, the Coulomb force
becomes relatively more important and the system effectively behaves
as if $T$ decreased: the polymer configuration becomes increasingly
aligned.

In the variational approach, as discussed above, the high temperature
regime is characterized by a purely fluctuating solution, reflecting
the Brownian nature of the chain at $T \rightarrow \infty$. Such a
solution survives as a local minimum also at lower $T$, where however
also a rigid solution exists. Below a certain critical temperature
$T_c$, the latter gives the global minimum, indicating a first order
phase transition. This is probably an artefact of the variational
approach - in the MC simulations the system shows no evidence of
possessing a phase transition (see section 5). The rigid solution
mirrors the ordered elongated structure of the polymer at low $T$.

With these qualitative arguments in mind, we turn to a more detailed
investigation of the behaviour of the polyelectrolyte in the high and
low $T$ limits, together with an evaluation of the corresponding
variational results.  The unscreened and screened cases will be
treated separately.

\subsection{The Unscreened Coulomb case}

\subsubsection{High Temperature}

In the high $T$ limit, the variational results can be expanded in
$1/T$ (see Appendix D). Thus, for the expectation value of the
Gaussian energy $E_G$, the first two terms of the expansion yield,
\beq
	\langle E_G\rangle = 3(N-1)T/2 + 1/\sqrt{2 \pi T}
	\sum_{k=1}^{N-1} \frac{N-k}{\sqrt{k}}
\eeq
which agrees with the exact result to the order shown; the first discrepancy
occurs in the $O(T^{-2})$ term, as is in fact true for any quadratic expectation
value $\langle \ri \rj \rangle$. By the virial identity, this also holds for the
Coulomb energy $\langle E_C\rangle$ and for the total energy as well.

For the individual bond lengths, the high $T$ result can be written
\beq
	\langle \ri^2 \rangle \approx 3T + 4\sqrt{2/\pi T} \left
	( \sqrt{i} + \sqrt{N-i} - \sqrt{N} \right )
\eeq
where the last term is obtained from a continuum approximation, valid
for large $N$.

\subsubsection{Low Temperature}

In the low $T$ limit (see Appendix D), the exact result for the total
internal energy, expanded in powers of $T$, is,
\beq
\label{E_0}
	\langle E\rangle = E_0 + (3N-5)T/2 + O(T^2)
\eeq
where $E_0$ is the minimum energy at $T = 0$ and $3N-5$ is the number
of degrees of freedom, modified for the spherical symmetry ($\ss
-2$).  It turns out that with the above first order low $T$
correction, the energy, and thus also $E_G$, $E_C$ and $r_{mm}$ (see
below), are quite well approximated for the sizes and temperatures
considered in this paper.

At low $T$, the variational free energy is minimized by the rigid
solution, for which the corresponding expansion is,
\beq
	\langle E\rangle_V = E_0 + 3(N-1)T/2 + O(T^2)
\eeq
where the first term is the same as in eq. (\ref{E_0}), while the
second term is qualitatively correct for large $N$.

Yet another low temperature expansion results from the purely
fluctuating variational solution and it gives,
\beq
	\langle E\rangle_V = (6/\pi)^{1/3} E_0 + 3(N-2)T/2 + O(T^2)
\eeq
which shows a 24\% discrepancy in the first term. The same factor,
$(6/\pi)^{1/3}$, results for any quadratic expectation value and for
any single term in $\langle E_C\rangle$ in this limit. For r.m.s.
distances, like the monomer-monomer distance $r_{mm}$,
\beq
\label{rm}
	r_{mm} = \left \langle \frac{1}{N-1}\sum_i \ri^2
	\right \rangle^{1/2}
\eeq
or the end-end distance $r_{ee}$,
\beq
\label{re}
	r_{ee} = \left \langle \left ( \sum_i \ri \right )^2 \right
	\rangle^{1/2}
\eeq
this corresponds to an error of 11\%.

With the variational approach thus satisfactory in both temperature
limits one can hope to find it a reasonable approximation also at
finite temperatures, as will indeed be borne out in section 5.

\subsubsection{Zero Temperature}

Having low $T$ expansions under control in terms of the $T=0$
configurational properties, the latter remain to be calculated.  They
are given by the minimum energy configuration, which is aligned,
\beq
	\ri = b_i \hat{\mbox{\bf n}}
\eeq
and unique (up to global translations and rotations).

The bond lengths $b_i>0$ satisfy the equation
\beq
	b_i = \sumsigi \frac{1}{b_{\sig}^2}
\eeq
where $b_{\sig} = \sumjsig b_j$ is the length of the subchain
$\sig$. This equation cannot be solved analytically (except for very
small $N$), but a fair large-$N$ approximation can be obtained (see
Appendix E for details).  This gives for a distinct monomer-monomer
bond the result
\beq
\label{riilog}
	\langle \ri^2 \rangle^{1/2}_{T=0} \equiv b_i \approx \left [
	\log \left ( \mbox{const} \; \frac{i(N-i)}{N} \right )
	\right ] ^{1/3}
\eeq
As a consequence, we find that at zero temperature the average
bond-length should scale logarithmically with $N$,
\beq
\label{rmmlog}
	r_{mm} \propto (logN)^{1/3}
\eeq
For the end-end separation this implies
\beq
\label{reelog}
	r_{ee} \propto N(\log N)^{1/3}
\eeq
This result is interesting, since it predicts a scaling faster than
$N$; the extra logarithmic factor comes from the stretching of the 
harmonic bonds.

The results above might seem as rather academic and of little
practical impact.  However, the low temperature expressions turn out
to be quite accurate when compared with MC results. In other words
ordinary room temperature and aqueous solution corresponds to a
surprisingly "low" temperature. This indicates that the low
temperature expansion might be a good starting point for further work
in polyelectrolyte solutions.

\subsection{The Screened Coulomb case}

At high $T$, an analysis similar to the one carried out for the pure Coulomb
chain, can be done for the screened chain (see Appendix D). Also there, the
variational approximations to quadratic expectation values turn out correct to
next-to-leading order; the same holds for $E$ and $E_G$, while $E_C$ (which is
one order down) is correct to leading order.

Also at low $T$, the results remain essentially the same for the
screened chain. Thus, the rigid solution gives correct energies to
lowest order in $T$, while the purely fluctuating solution does not.

\section{Monte Carlo Simulation Techniques}

For the numerical evaluation of the variational approach, the results
were compared to those from MC simulations, which were performed in
the canonical ensemble with the traditional Metropolis algorithm
\cite{metr}. For short chains this is a straightforward procedure, but
for linear chains consisting of more than about 100 monomers
convergence problems appear. Typically for a chain of 40 monomers four
million moves/monomer were required in order for the statistical
fluctuations in the end-end separation to be less than one per
cent. The energy terms and local conformational properties like
monomer-monomer separations converged much faster. The addition of
salt, i.e. the use of a screened potential, improves the convergence
characteristics, but on the other hand its evaluation is more
time-consuming than the pure inverse square root. Careful coding, with
table look-ups for the inverse square root routine and, in particular,
for the screened Coulomb potential, turned out to be more rewarding,
reducing the computation time by almost a factor of three. 

In order to treat longer chains with reasonable statistics we are
forced to use more efficient algorithms like the pivot algorithm, first
described in refs. \cite{lal,mad}, with a high efficiency for
linear chains on a lattice with short range interactions. Recently it
has also been used successfully \cite{bish} for off-lattice
simulations of a single polymer chain. It could be argued that the
pivot algorithm should be even more efficient for chains with long
range repulsive interactions like a charged polymer. The form of the
pivot procedure used in this work can be described as a two step
process, consisting of a random translation followed by a random
rotation or vice versa.  For a polymer chain with fixed bond lengths
only random rotations will be used. The procedure is as follows:
choose a monomer $i$ and apply the same random translation to monomers
$i+1$ to $N$. Then choose an axis at random and perform a random
rotation of monomers $i+1$ to $N$ around this axis.  Evaluate the
interaction energy between monomers $1$ to $i$, and $i+1$ to $N$. This
is a quadratic process in contrast to the single move algorithm, which
only requires the evaluation of $N$ pair interactions/move. Finally a
Metropolis energy criterion is used to test for rejection or
acceptance of the new configuration. We find that with a maximal
random displacement of the order of 5-10 \AA \ and a maximal random
rotation of $\pi$, we reject approximately 50 \% of the attempted
moves. Typically, we generate $10^3$-$10^4$ passes (one pass = one 
attempted move/monomer) resulting in
a statistical uncertainty in the end-end separation of approximately
one per cent. The uncertainty in the average monomer-monomer
separation and in the Coulomb and Gaussian energies is much
less. Local averages, however, like the $i$th bond length, may have
larger uncertainties, something that is also discussed by Madras and
Sokal \cite{mad}. The pivot algorithm seems to be superior to the
traditional single monomer procedure described initially already for
chains with $N>20$. We also have found that restricting the procedure
to translational moves still makes it superior to the traditional
algorithm. The pivot algorithm makes it feasible to simulate chains 
with more than one thousand interacting monomers.
The major drawback with the pivot algorithm seems to be its
limitations to linear chains or at least chains with simple
topologies.

Without excessive fine tuning of the translational and rotational
displacement parameters, we find that the computational cost grows as
$N^3$. This power results from $N^2$ for each sweep of monomer moves, and an
additional factor $N$ from autocorrelations in quantities like end-end
separations.

Another efficient algorithm has recently been developed in
ref. \cite{irb}.  By identifying the slow modes in a Fourier analysis,
one is able to use different random step lengths for different modes.
This technique seems to be as efficient as the pivot algorithm and we
have used it to check the accuracy of our simulations. For all
cases investigated we obtain, within the statistical uncertainties,
identical averages. The same is true for shorter chains where we also
can use the original single monomer algorithm as a further test.

\section{Numerical Results}

\begin{figure}[htb]
\psfig{figure=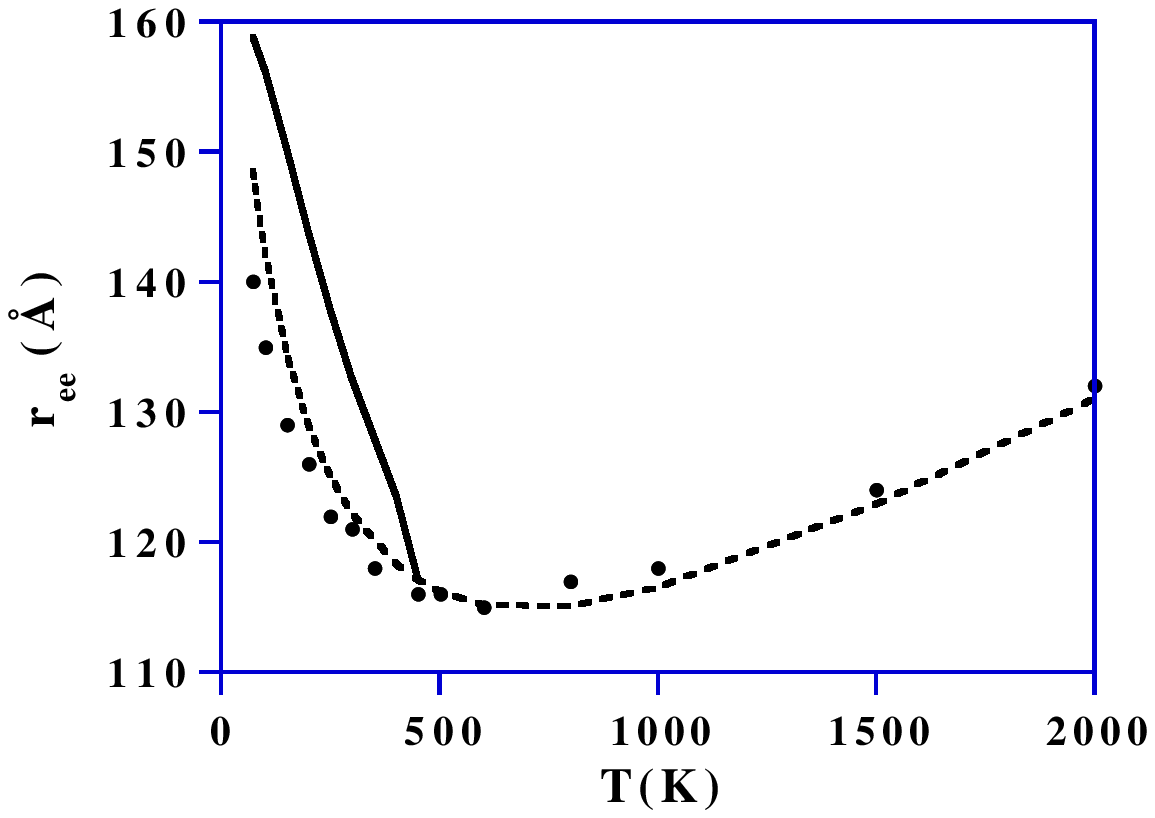,width=4in,height=3in}
\psfig{figure=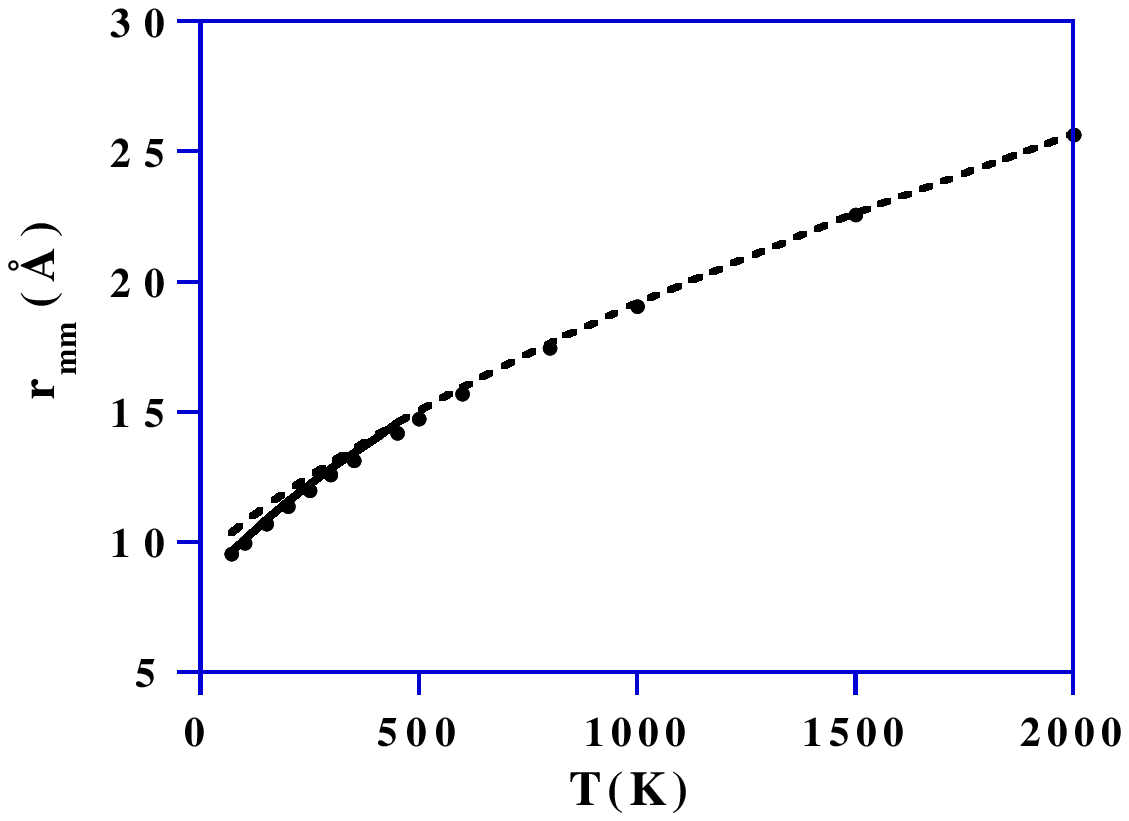,width=4in,height=3in}
\caption{{\bf (a)} $r_{ee}$ and {\bf (b)} $r_{mm}$ as functions of $T$ for
an unscreened chain with $N=20$. Filled circles represent MC data, and
solid and dashed lines variational results, with $a \neq 0$ and $a = 0$,
respectively.}
\label{r_Tab}
\end{figure}

\begin{figure}[htb]
\psfig{figure=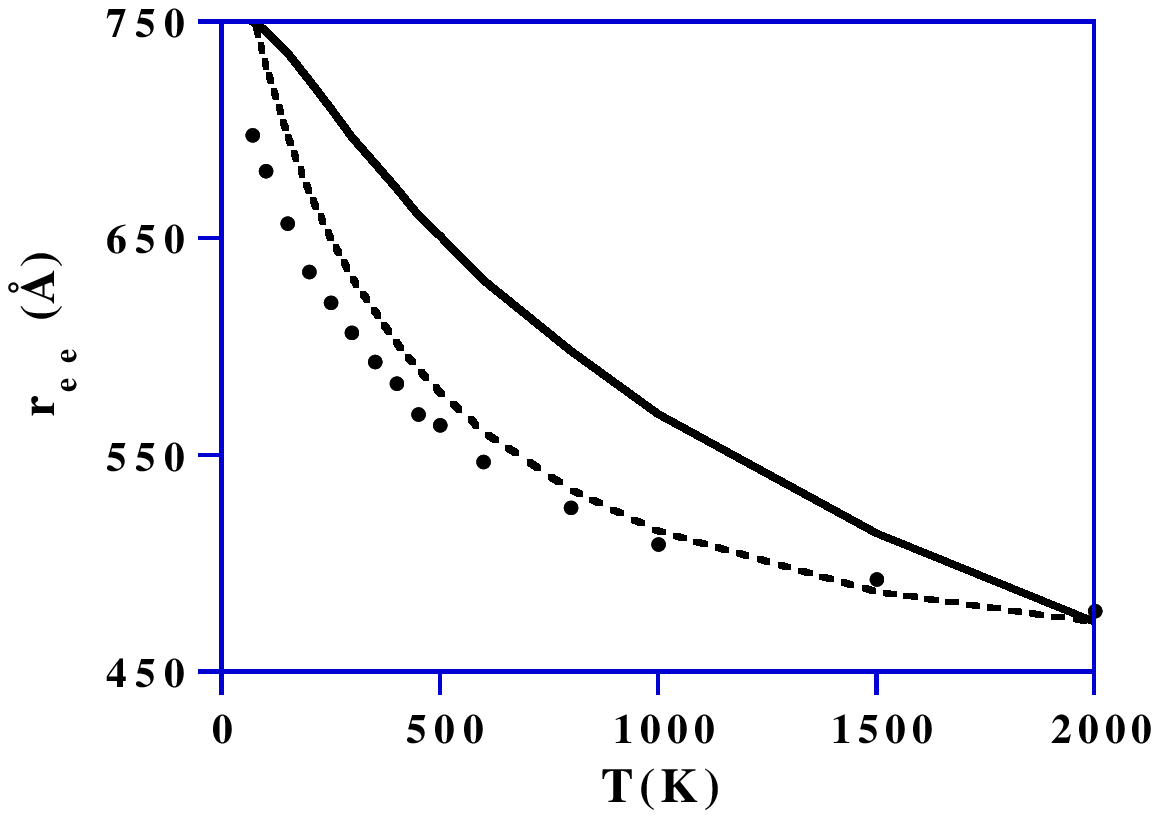,width=4in,height=3in}
\psfig{figure=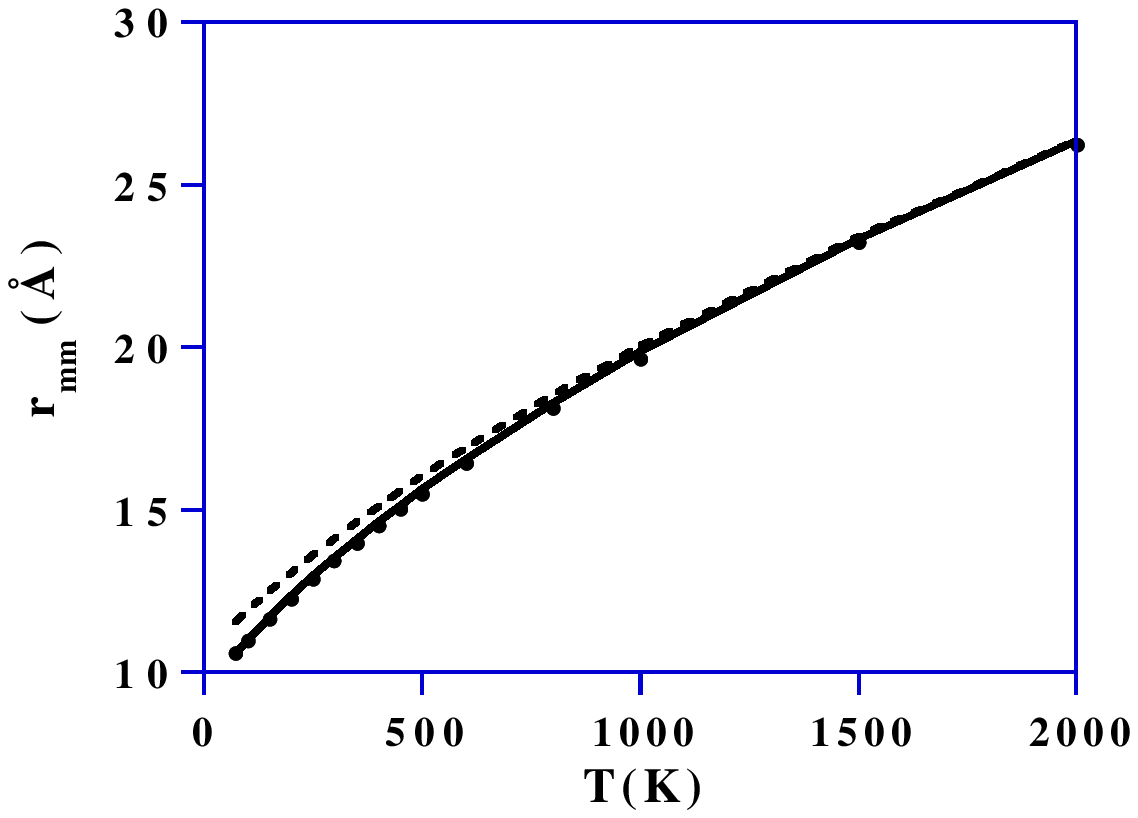,width=4in,height=3in}
\caption{{\bf (a)} $r_{ee}$ and {\bf (b)} $r_{mm}$ as functions of $T$ for
an unscreened chain with $N=80$. Same notation as in fig.
\protect\ref{r_Tab}.}
\label{r_Tcd}
\end{figure}

The superiority of the purely fluctuating variational solution over the
rigid one, as discussed in section 5, is illustrated in figs. \ref{r_Tab}
and \ref{r_Tcd}, where the variational results for $r_{ee}$ and $r_{mm}$
are compared to MC data for $N=20$ and 80.

The remainder of this section will be devoted to comparisons of the
variational approach to MC simulation results focusing on (1)
energies, in particular the free energy and (2) various
configurational measures.  The $\ai = 0$ variational solution will be
consistently used.

\subsection{Free and Internal Energies}

In table \ref{t_E} the variational results for internal Coulombic and
Gaussian energies are compared with the MC results; the relative
deviations are seen to increase both with increasing $c_s$ and with
increasing $N$.
\begin{table}[htb]
\begin{center}
\begin{tabular}{|ccr|c|c|c|c|c|c|}
\hline
& & N & 20 & 40 & 80 & 160 & 320 & 512 \\
\hline \hline
&$c_s$=0.0 &       $E_C$ (V) & 6.20 & 7.58 & 8.80 & 9.94 & 11.0 & 11.7 \\
&          &            (MC) & 5.25 & 6.30 & 7.28 & 8.16 & 8.99 & 9.53 \\
\hline
&          &       $E_G$ (V) & 6.65 & 7.40 & 8.08 & 8.66 & 9.20 & 9.54 \\
&          &            (MC) & 6.25 & 6.78 & 7.31 & 7.79 & 8.23 & 8.47 \\
\hline \hline
&$c_s$=0.01 &      $E_C$ (V) & 3.55 & 3.80 & 3.95 & 4.02 & 4.05 & 4.07 \\
&          &            (MC) & 2.70 & 2.83 & 2.89 & 2.91 & 2.93 & 2.93 \\
\hline
&          &      $E_G$ (V)  & 6.20 & 6.63 & 6.88 & 7.00 & 7.07 & 7.10 \\
&          &            (MC) & 5.70 & 6.00 & 6.15 & 6.22 & 6.26 & 6.27 \\
\hline \hline
&$c_s$=0.1 &      $E_C$ (V)  & 1.90 & 2.03 & 2.16 & 2.19 & 2.15 & 2.15 \\
&          &            (MC) & 1.35 & 1.38 & 1.40 & 1.41 & 1.42 & 1.42 \\
\hline
&          &      $E_G$ (V)  & 5.40 & 5.68 & 5.86 & 5.94 & 5.93 & 5.94 \\
&          &            (MC) & 5.00 & 5.15 & 5.26 & 5.29 & 5.32 & 5.32 \\
\hline \hline
&$c_s$=1.0 &      $E_C$ (V)  & 0.65 & 0.70 & 0.74 & 0.75 & 0.76 & 0.76 \\
&          &            (MC) & 0.40 & 0.40 & 0.41 & 0.43 & 0.42 & 0.42 \\
\hline
&          &      $E_G$ (V)  & 4.35 & 4.53 & 4.61 & 4.67 & 4.69 & 4.70 \\
&          &            (MC) & 4.10 & 4.25 & 4.29 & 4.33 & 4.37 & 4.37 \\
\hline \hline
\end{tabular}
\end{center}
\caption{Average internal Coulombic and Gaussian energies per monomer
in $kJ/mol \cdot monomer$ for unscreened and screened Coulomb potential. 
MC and V stands for Monte Carlo and variational calculations respectively. 
The salt concentration, $c_s$, is given in molar (M).}
\label{t_E}
\end{table}
For fixed $c_s$ the deviations seem to converge to constant values at
large $N$. Hence it should be possible to extract "asymptotic"
correction factors from comparisons at moderate $N$, do a variational
calculation for a very large $N$, and predict what an MC calculation would
give.

\begin{figure}[htb]
\psfig{figure=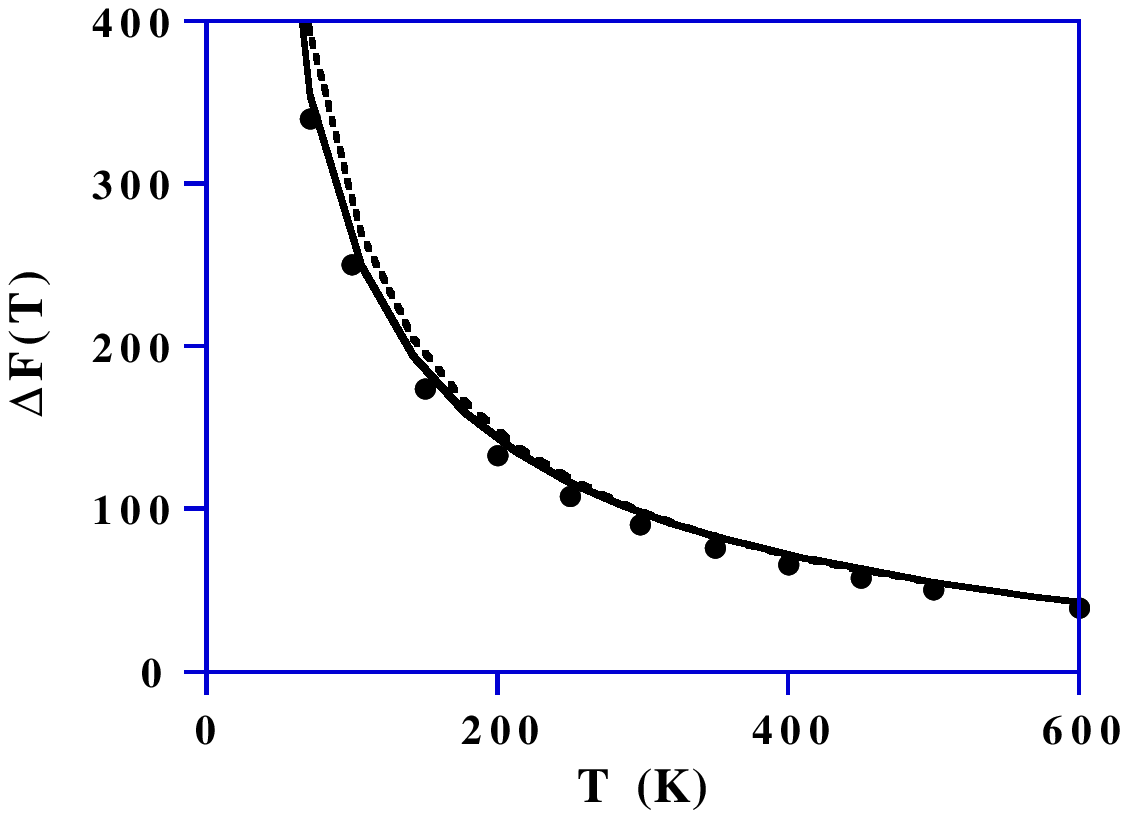,width=4in,height=3in}
\psfig{figure=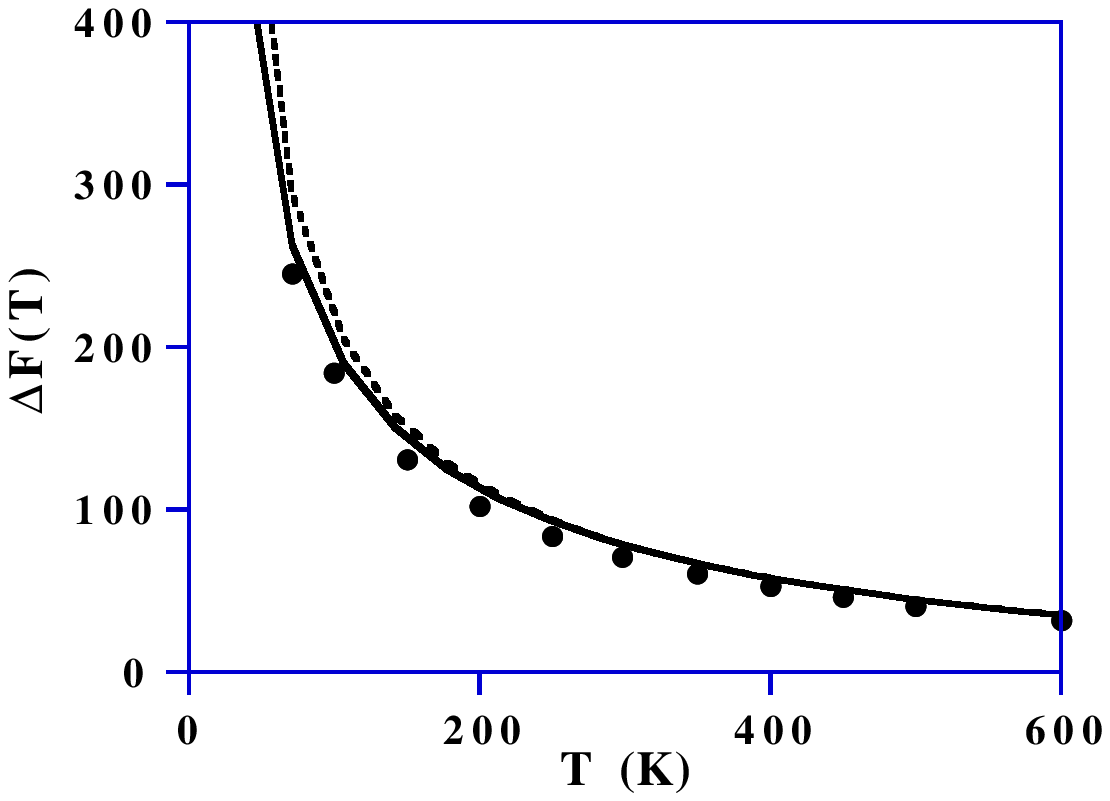,width=4in,height=3in}
\caption{ The excess free energy $\Delta F(T)$ (see eq.
(\protect\ref{F_ex})) from the MC data (filled circles) and from the
variational approach (solid line with $a \neq 0$ and dotted line with $a =
0$) for {\bf (a)} $c_s=0.0 M$ and {\bf (b)} $c_s =0.1 M$ respectively
($N$=10).}
\label{DF}
\end{figure}

A strong advantage of the variational approach is the direct access to the
free energy, which is much more difficult to obtain in a MC simulation,
requiring a cumbersome integration procedure. In order to evaluate the
variational results we nevertheless attempt to estimate F(T) from MC data
for $N=20$ using the following procedure. In dimensionless units one has
\beq
	\frac{d(F/T)}{dT} = - \frac{d}{dT} \log \int e^{-E/T} dx =
	-\frac{1}{T^2} \langle E \rangle
\eeq
Thus, we can define an {\it excess free energy} with respect to some
reference temperature $T_r$ as
\beq
\label{F_ex}
	\Delta F(T) = F(T)/T - F(T_r)/T_r = -\int_{T_r}^T \langle E
	\rangle dT/T^2
\eeq
which is then 
accessible in MC by a temperature integration of $\langle E \rangle$.  In
fig. \ref{DF} the excess free energy is shown as a function of $T$, for an
$N=20$ chain with $T_r$ corresponding to 1422 K.
As can be seen the variational solutions for $F(T)$ reproduce the
extracted MC values very well in a wide temperature interval.
Comparisons for larger $N$ are not feasible due to the indirect
cumbersome MC extraction procedure.

We remark that one possibly efficient alternative route to obtain free
energies for different degrees of screening at fixed $T$ in MC, would be
to perform an integration in the Debye screening length and calculate the
incremental excess free energy for a change $\Delta \kappa$. The advantage
of such a procedure would be that every point along the integration path
corresponds to a physically realistic situation, while the completely 
screened chain would serve as a reference state.

\subsection{The End-End Separation}

\begin{figure}[htb]
\psfig{figure=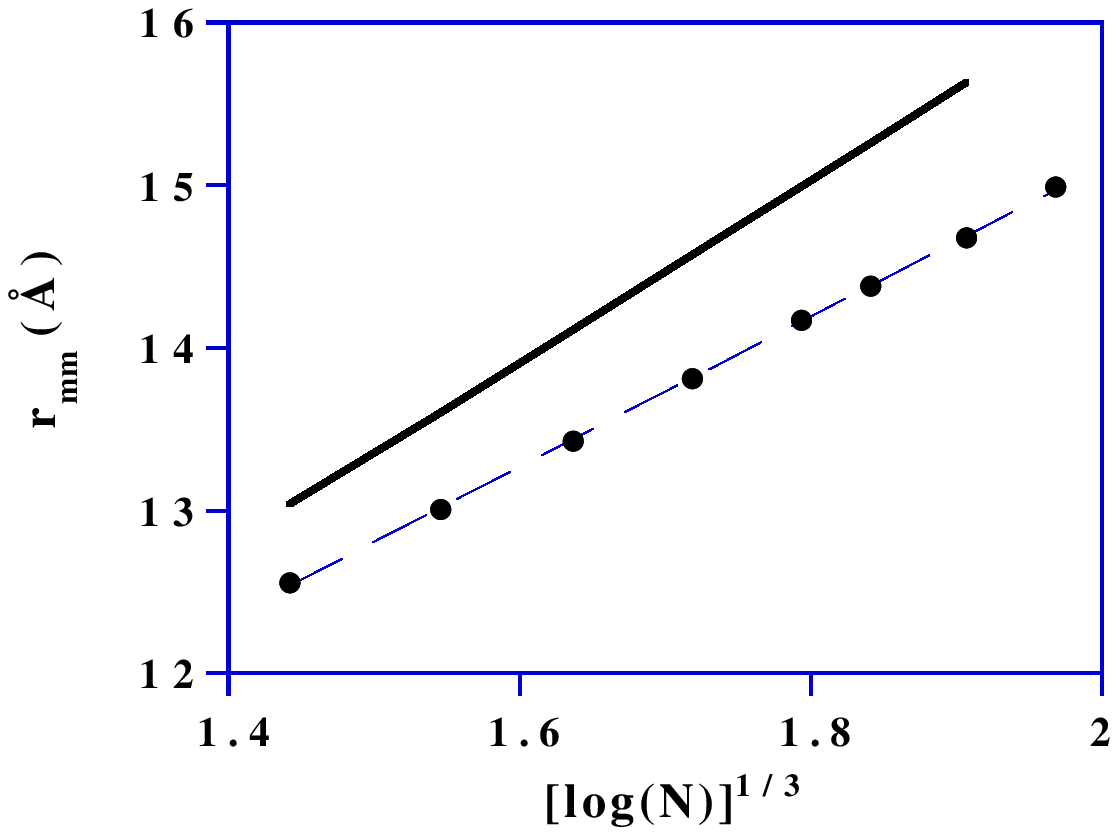,width=4in,height=3in}
\psfig{figure=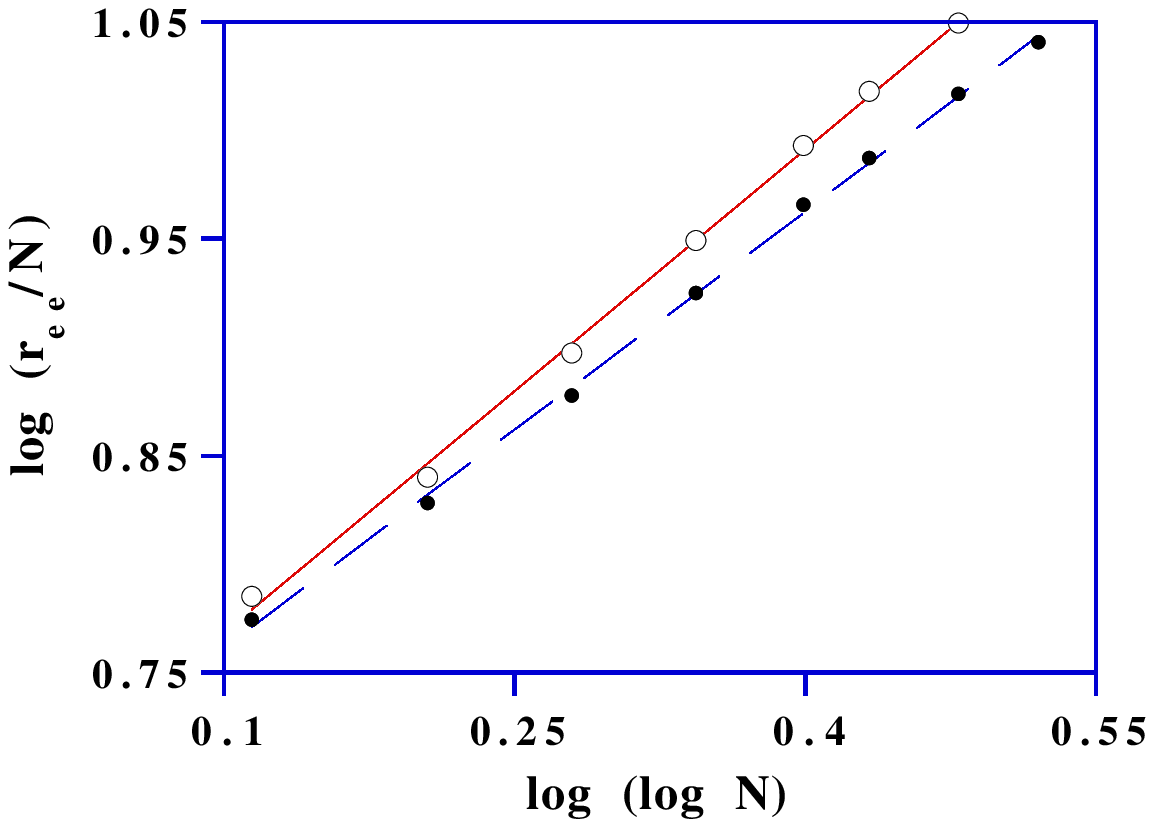,width=4in,height=3in}
\caption{ {\bf (a)} $r_{mm}$ as a function of $(\log N)^{1/3}$. Filled circles 
represent MC data and solid line the variational results. The dashed line is 
a linear fit to the MC data. {\bf (b)}  $\log r_{ee}$ as a function of 
$\log (\log N)$.  Filled and open circles represent MC data and 
variational results respectively. The lines are linear fits.}
\label{ree_N_0}
\end{figure}

The end-end separation is a critical measure of the accuracy,
being a global quantity with contributions from all bond-bond
correlations.  Unfortunately, it also turns out to be a complicated
quantity from the convergence point of view in standard MC simulations, which
means that it will have larger uncertainties than for example the average
monomer-monomer separation -- this is particularly true for the pure
Coulomb chain. Table \ref{t_r_0} contains a comparison between variational
and simulation results for $r_{mm}$ and $r_{ee}$ using the unscreened
Coulomb potential. The result from the variational approach is impressive
-- the maximal deviation, 7.8\% for $N=1024$, from the MC results is well
below the 11\% bound discussed above. (Due to memory limitations on the 
local workstation $N=2048$ has not been pursued with the variational 
approach.) This result can be compared to the
results from a (spherical) mean field approach \cite{wood1}, which for the
same system shows a deviation of 20\% already for $N=100$.
\begin{table}
\begin{center}
\begin{tabular}{|cc|c|c|c|c|c|c|c|c|}
\hline
& $N$ & 20 & 40 & 80 & 160 & 320 & 512 & 1024 & 2048 \\
\hline \hline
$r_{mm}$ & V         & 13.04  & 13.60 & 14.11 & 14.57 & 14.99 & 15.26 & 15.63 &       \\
& MC                 & 12.56  & 13.01 & 13.43 & 13.81 & 14.17 & 14.38 & 14.68 & 14.99 \\
& diff.              & 3.8\%  & 4.5\% & 5.1\% & 5.5\% & 5.8\% & 6.1\% & 6.5\% &       \\
\hline
$r_{ee}$ & V         & 122    & 277   & 632   &  1425 &  3152 & 5340  & 11478 &       \\
& MC                 & 119    & 269   & 606   &  1347 &  2958 & 4985  & 10651 & 22507 \\
& diff.              & 2.5\%  & 3.9\% & 4.3\% & 5.8\% & 6.6\% & 7.1\% & 7.8\% &       \\
\hline \hline
\end{tabular}
\end{center}
\caption{$ r_{mm}$ and $r_{ee}$ in \AA \ for unscreened Coulomb
potential ($c_s=0.0M$) as computed with the variational (V) and Monte
Carlo (MC) methods. The errors originating from the MC runs
are estimated to be $O$(0.2\%).}
\label{t_r_0}
\end{table}
In section 4, $r_{ee}/N$ and $r_{mm}$ were conjectured to vary linearly
with $(\log N)^{1/3}$ for large $N$ at zero temperature (cf. eqs. 
(\ref{rmmlog}, \ref{reelog})). In fig. \ref{ree_N_0}, this linear
dependence is tested on both MC and variational data at room temperature
(298K), with a surprisingly good result. This indicates that room
temperatures can be considered low for a reasonably long Coulomb chain
with the chosen parameters. It is also clearly seen how the variational
approximation to $r_{ee}$ exceeds the MC values; asymptotically we expect
an 11\% discrepancy, as discussed above.

The relative errors for $r_{ee}/N$ and $r_{mm}$ in table \ref{t_r_0}
may be used for a numerical estimate of the asymptotic error. Assuming that
the relative error decays to the final value like $(\log N)^{-1/3}$ we
estimate the asymptotic errors in $r_{ee}$ and $r_{mm}$ to be 10 and 13 \%,
respectively, in good agreement with the expected 11 \%.

The difference in slope for the variational and simulated $r_{ee}/N$
in fig. \ref{ree_N_0}b is a consequence of the finite number of monomers.
Numerically we find, at the present temperature, that $N=$ 1000-2000 is not
enough in order to reach the asymptotic regime -- the slopes in fig.
\ref{ree_N_0}b are larger than 1/3. By lowering $T$, however, we expect both
slopes to approach this limiting value.
%
%
%

The conjectured zero-temperature scaling results seem to contradict
the results by deGennes et al. \cite{gen1} and Baumg\"artner
\cite{baum} that $r_{ee}$ should scale linearly with $N$ for an
unscreened polyelectrolyte. However, the latter result is valid only
if the monomer-monomer bonds are rigid, but with elastic bonds as in
the present model, eq. (\ref{E}), swelling is possible and $r_{ee}/N$
increases slowly with $N$ due to the long-range Coulomb repulsion.

%
%
%
%
Returning to the simulations by Baumg\"artner \cite{baum}, we note that 
his effective temperature is almost a factor of ten larger than in the 
present study. Such a high temperature means that the chain behaviour 
is essentially brownian in character making it numerically difficult to 
detect the electrostatic expansion of the chain in a traditional MC 
simulation. The rigid monomer-monomer bonds used by Baumg\"artner also 
precludes the extra expansion predicted by eqs. (\ref{rmmlog},\ref{reelog}).

Table \ref{t_r_cs} contains the same quantities as table \ref{t_r_0}, but
for a screened Coulomb potential. The agreement detoriates, when a small
amount of salt is added. The screening reduces the Coulomb repulsion,
which in a sense is the hard part in our variational calculation, and one
would naively expect an improved accuracy with a decreased interaction. The
opposite result is found and the discrepancy in the end-end separation
becomes as large as about 40\% with 10 mM of salt and $N=160$. At
sufficiently high salt concentration the agreement improves again, as it
should, with the Coulomb repulsion completely screened.  With a strong
screening the Coulomb potential will have a short range, and for
sufficiently large $N$ the chain will be Brownian. This is also reflected
in table \ref{t_r_cs}, where the agreement for $r_{ee}$ is worst for intermediate
chain lengths and improve again when $N$ increases.  The large
discrepancy seen e.g. for $c_s$=0.01 M and $N$=160 might seem surprising,
considering the excellent agreement found for the pure Coulomb chain, but
it reflects the difficulty to properly emulate a short-range potential
with a harmonic effective energy \cite{mez1}.
\begin{table}
\begin{center}
\begin{tabular}{|ccc|c|c|c|c|c|c|}
\hline
& & $N$ & 20 & 40 & 80 & 160 & 320 & 512 \\
\hline \hline
$c_s$=0.01 & $r_{mm}$ & V  & 12.60  & 12.87 & 13.02 & 13.10 & 13.14 & 13.16 \\
& & MC                     & 12.09  & 12.24 & 12.31 & 12.34 & 12.36 & 12.37 \\
& & diff.                  & 4.2\%  & 5.1\% & 5.8\% & 6.1\% & 6.3\% & 6.4\% \\
\hline
& $r_{ee}$ & V             & 104    & 201   & 377   &  680  & 1188  & 1710 \\
& & MC                     & 99.6   & 183   & 317   &  521  &  825  & 1110 \\
& & diff.                  & 4.4\%  & 9.8\% & 19\%  & 31\%  & 44\%  & 54\% \\
\hline \hline
$c_s$=0.1 & $r_{mm}$ & V   & 11.77  & 11.90 & 11.97 & 12.01 & 12.04 & 12.04 \\
& & MC                     & 11.30  & 11.35 & 11.39 & 11.38 & 11.40 & 11.40 \\
& & diff.                  & 4.2\%  & 4.8\% & 5.1\% & 5.5\% & 5.6\% & 5.6\% \\
\hline
& $r_{ee}$ & V             & 78.2   & 136   & 231   &  387  & 640   & 895 \\
& & MC                     & 72.9   & 120   & 192   &  301  & 459   & 622  \\
& & diff.                  & 7.3\%  & 13\%  & 20\%  & 29\%  & 39\%  & 44\% \\
\hline \hline
$c_s$=1.0   & $r_{mm}$ & V & 10.57  & 10.69 & 10.69 & 10.69 & 10.70 & 10.70 \\
& & MC                     & 10.27  & 10.29 & 10.29 & 10.30 & 10.34 & 10.32 \\
& & diff.                  & 2.9\%  & 3.9\% & 3.9\% & 3.8\% & 3.5\% & 3.7\% \\
\hline
& $r_{ee}$ & V             & 55.0   & 86.9  & 137   &  217  &  343  & 468  \\
& & MC                     & 52.1   & 79.5  & 122   &  182  &  283  & 364  \\
& & diff.                  & 5.6\%  & 9.3\% & 12\%  &  19\% &  21\% & 29\% \\
\hline \hline
\end{tabular}
\end{center}
\caption{$ r_{mm}$ and $r_{ee}$ in \AA \ for the screened Coulomb potential
($c_s$=0.01M,$c_s$=0.1M and $c_s$=1.0M respectively) as computed with the
variational (V) and Monte Carlo (MC) methods. The errors originating from
the MC runs are estimated to be $O$(0.1\%)}
\label{t_r_cs}
\end{table}

\subsection{Monomer-Monomer Separations}

\begin{figure}[htb]
\psfig{figure=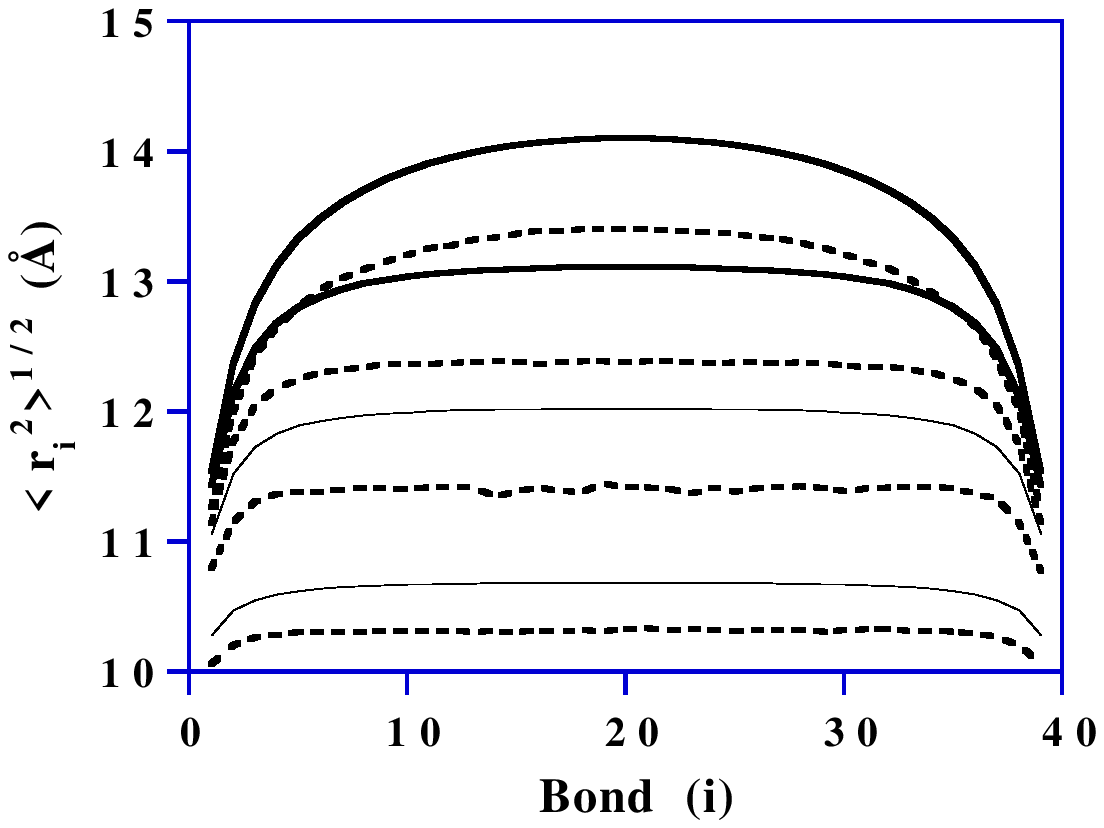,width=4in,height=3in}
\caption{Bond lengths $\langle \ri^2 \rangle^{1/2}$ along
an $N=40$ chain.  Salt concentrations $c_s$ from top to bottom are 0 M,
0.01M, 0.1M and 1 M respectively. Solid line represents variational
results and dashed line Monte Carlo data.}
\label{rii_cs}
\end{figure}

The variational results for the average monomer-monomer separation
$r_{mm}$ is in excellent agreement with MC results for both the unscreened
and screened cases - the largest error seen is of the order of 5\% (see
tables \ref{t_r_0} and \ref{t_r_cs}). As for $r_{ee}$ the variational
estimate is always larger than the MC value. This is also true for any
single monomer-monomer separation $\langle \ri^2 \rangle^{1/2}$ as can be
seen in fig. \ref{rii_cs}.
The shape of the curves are correctly reproduced by the variational
solution and the largest discrepancy is not unexpectedly found in the
middle of the chain, which may be explained by a stronger accumulated
electrostatic repulsion there. These results can be compared to the mean
field solution of ref. \cite{gran}, in which gradually more and more pair
interactions were treated explicitly and withdrawn from the mean field.
It was found that only after the inclusion of next-nearest neighbours,
leading to lengthy numerical calculations, the curve shape became
qualitatively correct.

We discussed above the zero-temperature scaling for an unscreened chain,
$r_{mm} \propto (\log N)^{1/3}$.  When salt is added $r_{mm}$ increases
much more slowly with chain length and for $c_s$=1 M, when $r_{mm} \gg
\kappa ^{-1}$, it is essentially independent of chain length. One also 
notes that the individual bond lengths $\langle r_i^2 \rangle^{1/2}$ do 
not vary in the central part of the chain when salt is present 
(see fig. \ref{rii_cs}).

\begin{figure}[htb]
\psfig{figure=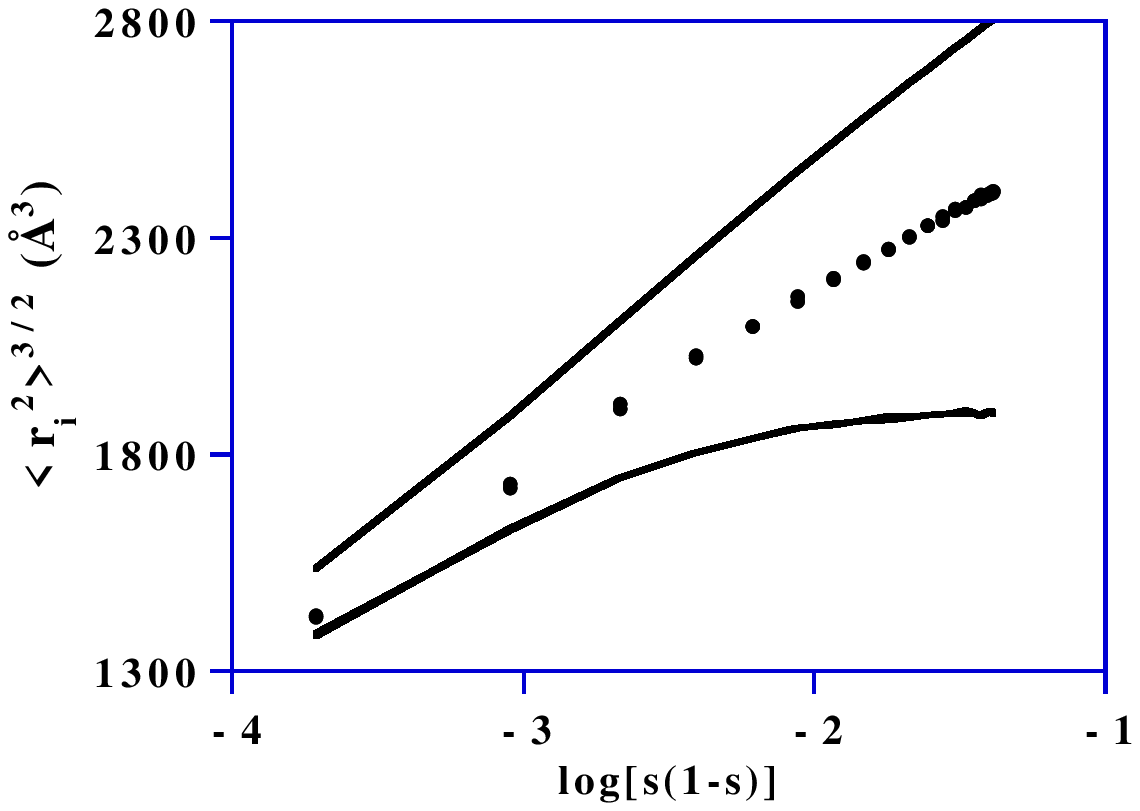,width=4in,height=3in}
\psfig{figure=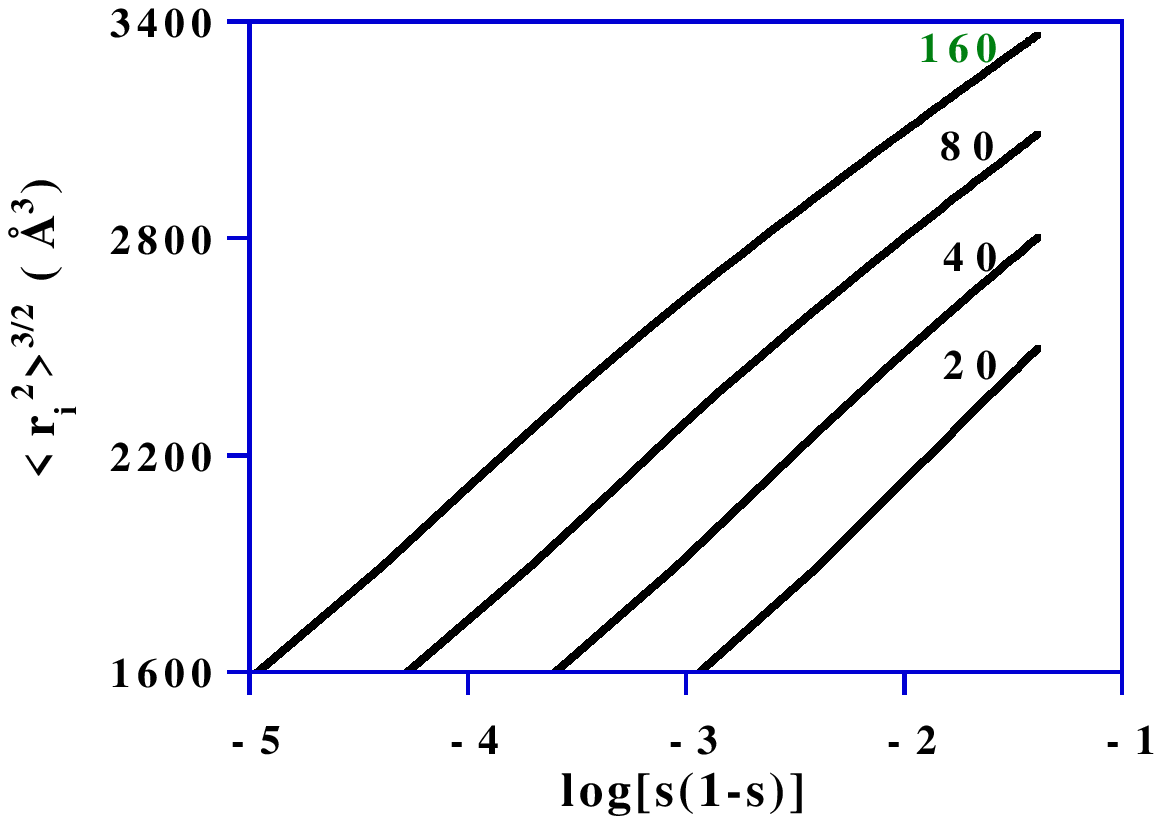,width=4in,height=3in}
\caption{The cubed bond length $\langle \ri^2 \rangle^{3/2}$ as a
function of $\log(s(1-s))$, with $s=i/N$ the relative monomer
position.  {\bf (a)} Upper line shows variational results and symbols
MC data for $N=40$, no screening. The zero temperature scaling
corresponds to a straight line. The lower line shows MC data for a
screened chain ($c_s=0.01$ M) for comparison.  {\bf (b)} Variational
results for unscreened chains of varying size $N$.}
\label{figlog}
\end{figure}

Furthermore, for an unscreened chain, following eq. (\ref{riilog}), we
expect at $T=0$ the individual bond-lengths, raised to the power three, to
vary linearly with $u = \log [s(1-s)]$, where $s=i/N$. In fig.
\ref{figlog}a it is shown that this is approximately true, more so for the
variational solution than for the MC results.
The lower curve in fig. \ref{figlog}a, obtained with a screened Coulomb
potential, shows a qualitatively different behaviour. Fig. \ref{figlog}b
contains a similar graph for different $N$.

The $T=0$ scaling relation for individual bonds, eq. (\ref{riilog}), has the 
peculiar consequence that the length of a bond at the end of
the chain becomes independent of $N$. This can be seen
by rewriting the scaling relation as
\beq
\label{sc_o}
	\langle \ri^2 \rangle^{1/2} \approx \{ \log N + \log [ s(1-s) ]
	\}^{1/3} \approx [\log i]^{1/3}
\eeq
where the last expression holds for small $i$. Eq. (\ref{sc_o}) also 
holds for the MC results, where we find the first few bond lengths to 
be independent of $N$.

\subsection{Angular Correlations}

\begin{figure}[htb]
\psfig{figure=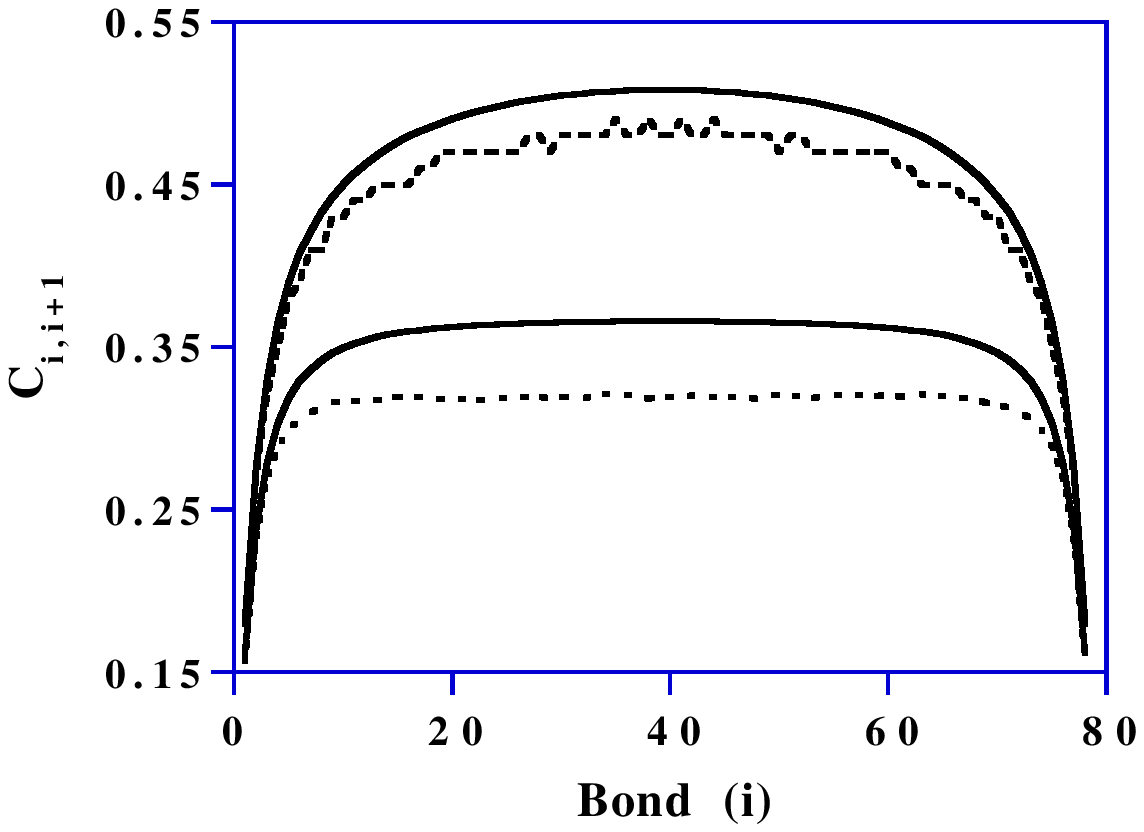,width=4in,height=3in}
\psfig{figure=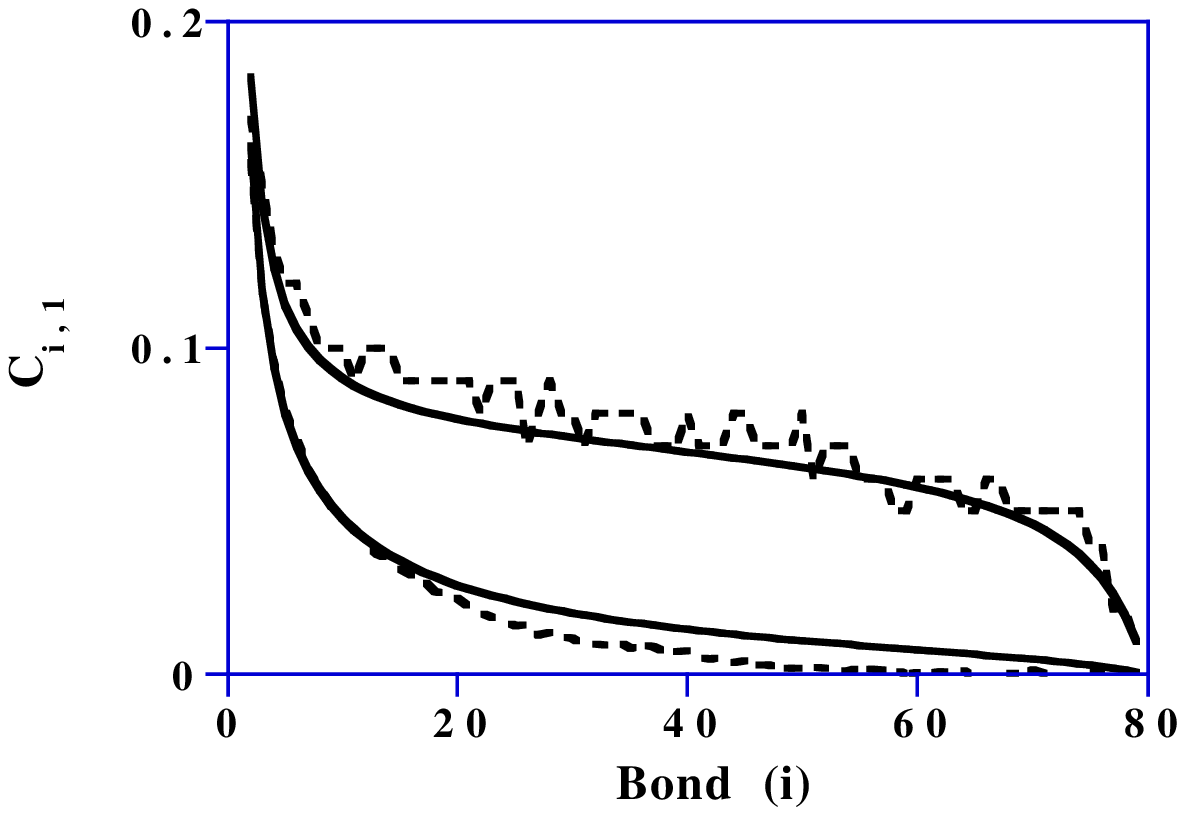,width=4in,height=3in}
\caption{Angular correlations {\bf (a)} between neighbouring bonds, 
and {\bf (b)}
between the first and successive bonds.  Solid lines represent variational
and dashed lines MC results. The upper pair of curves are for an
unscreened chain, while the lower pair is for $c_s$ = 0.01 M. ($N=80$.)}
\label{figang}
\end{figure}

In order to further test the variational solutions, we also have
calculated angular correlations between bonds,
\beq
	C_{i,j} = \frac{\langle \ri \cdot \rj \rangle}{\sqrt{\langle
	\ri^2\rangle \langle \rj^2 \rangle}}
\eeq
which roughly gives the average of the cosine between bonds.  In
fig. \ref{figang}a, variational and MC data is shown for the neighbor
correlation $C_{i,i+1}$. It is seen that the variational Ansatz
consistently overestimates the angle, more so in the presence of salt than
without.
Comparing figs. \ref{rii_cs} and \ref{figang}a, we find that the above
discussed discrepancy between the MC and variational results for the
end-end separation seems to be due to differences in angular correlations
as well as in bond lengths. For the unscreened case the two sources seem
to be of comparable magnitude, while for the screened case the angular
correlations seem to be larger and the main cause of discrepancy.

Fig. \ref{figang}b shows a more global angular correlation $C_{1,i}$
between the first and all successive bonds. The variational results for
this quantity are in excellent agreement with MC data, both in the
screened and in the unscreened chain.
%

\section{Summary and Outlook}

A deterministic variational scheme for discrete representations of polymer
chains has been presented, where the true bond and Coulombic potentials are
approximated with a trial isotropic harmonic energy. The variational
parameters obey matrix equations, for which a very effective iterative
solution scheme has been developed -- the computational demand is $N^3$.

The high and low T properties of the variational approach has been
analyzed with encouraging results. Also, the approach is shown to obey
the relevant virial identities.

In contrast to MC simulations, the free energy is directly accessible
with the variational method.

When confronting the results from the method with those from MC
simulations, very good agreement is found for configurational
quantities in the case of an unscreened Coulomb interaction (the
error is within 11 \%).

In the screened case the method does not reproduce the MC results equally 
well although the qualitative picture of conformational properties is
there. We attribute this problem to the difficulty for a Gaussian to emulate
short range interactions.

Recently, MC simulations were pursued for titrating Coulomb chains 
\cite{ull}. For such systems the Coulomb potential of eq. (\ref{E}) is 
modified to 
\beq
\label{E_c_s}
\frac{q^2}{4\pi \epsilon _r \epsilon_0}\sum_{i<j}
\frac{s_is_j}{|\tilde{\x}_{ij}|}
\eeq
where the binary variables $s_i$ are either $1$ or $0$ depending
whether monomer $i$ is charged or not. Thus minimizing E now also
includes a combinatorial problem -- deciding where the charges should
be located.  Variational techniques related to the ones used in this
paper have been successfully used in pure combinatorial optimization
problems \cite{pet1}, where again tedious stochastic procedures are
replaced by a set of deterministic equations. Along similar lines, the
approach of this paper can be modified to allow for a variational
treatment also of the titrating problem.

The variational approach is also directly applicable to more general
topologies -- bifurcations pose no problems.  Proteins could also be
treated in this way provided the traditional Lennard-Jones potentials are
replaced by forms that are less singular at the origin.

\section*{Acknowledgements:}

We wish to thank Anders Irb\"ack for providing us with some of the MC data
points.  One of us (BJ) wants to acknowledge many stimulating discussions
with R. Podgornik during the course of this work.

\appendix
\renewcommand{\thesection}{Appendix \Alph{section}.}
\renewcommand{\theequation}{\Alph{section}\arabic{equation}}

\newpage
\section{The Variational Approach -- Generalities}
\setcounter{equation}{0}

In this appendix we discuss the generic variational approach that is used
in this paper for the particular problem of a Coulomb chain.  We consider
a generic system, with dynamical variables $x$ in some multi-dimensional
state-space, and assume that a real energy function $E(x)$ is given.

For an arbitrary probability distribution $P(x)$ in an arbitrary state
space, the {\em free energy} with respect to an energy $E(x)$ is generally
defined as
\beq
	\hat{F} = \langle E \rangle - T S
\eeq
where $S$ is the {\em entropy},
\beq
	S = -\langle \log P \rangle
\eeq
and expectation values are defined with respect to $P$. Writing $P(x)$ as
\beq
	P(x) = \frac{1}{Z} \exp(-E_V(x)/T)
\eeq
with
\beq
	Z = \int dx \; \exp(-E_V(x)/T)
\eeq
the free energy can be written as
\beq
	\hat{F} = -T \log Z + \langle E - E_V \rangle
\eeq

Note that
\beqa
	\exp(-\hat{F}/T) & = & Z \exp \left \langle \frac{E_V - E}{T} \right
	\rangle \\ \nonumber & \leq & Z \left \langle \exp \left ( \frac{E_V -
	E}{T} \right ) \right \rangle = \int dx \; \exp(-E/T) \\ \nonumber & =
	& \exp(-\hat{F}/T)|_{E_V=E}
\eeqa
where the inequality is due to the convexity of the exponential
function. Thus, $\hat{F}$ is bounded from below by its value for $E_V
= E$, corresponding to the proper Boltzmann distribution, $P(x)
\propto \exp(-E(x)/T)$.

The variation of $\hat{F}$ due to a variation $\delta E_V$ is given by
\beqa
	\delta\hat{F} & = & -T \left ( \langle \delta E_V (E - E_V) \rangle -
	\langle \delta E_V \rangle \langle E - E_V \rangle \right ) \\
	\nonumber & \equiv & -T \langle \delta E_V (E - E_V) \rangle_C
\eeqa
where $\langle a b \rangle_C$ stands for the connected expectation value
(cumulant), $\langle a b \rangle - \langle a \rangle \langle b \rangle$.

The idea of the {\em variational approach} is to choose a suitable
simple Ansatz for the variational energy $E_V$, with a set of
adjustable parameters $\alpha_i$, $i = 1,\ldots,N_p$, the values of
which are to be chosen so as to minimize the variational free energy.
Demanding the vanishing of the variation of the free energy due to
variations in the parameters $\alpha_i$ then leads to the general
equations for an extremum:
\beq
\label{delF}
	\left \langle \frac{\partial E_V}{\partial \alpha_i} (E - E_V) \right
	\rangle^V_C = 0, \; \; i = 1, \ldots, N_p
\eeq
where $\langle \rangle^V$ denotes an expectation value based on the
variational Boltzmann distribution.  This determines the optimal
values of the parameters. Exact expectation values are then
approximated by the corresponding variational ones.
%

\newpage
\section{The Variational Free Energy for the Chain}
\setcounter{equation}{0}

In this appendix we derive the expressions for the variational free
energy (eqs. (\ref{F}, \ref{F_s})), for the unscreened as well as the
screened Coulomb chain, with or without translational parameters in
the variational Ansatz.

\subsection*{Unscreened Coulomb Chain}

For the specific case of the Coulomb chain of length $N$, the energy
amounts to
\beq
	E = \half \sum_i \ri^2 + \sumsig \frac{1}{r_{\sig}}
\eeq
where $\sig$ is a contiguous subchain.

\subsubsection*{Gaussian Parameters Only}

We consider first a pure Gaussian variational Boltzmann distribution,
corresponding to
\beq
\label{PG}
	E_V/T = \half \sum_{i,j} \Ginvij \, \ri \cdot \rj
\eeq
The parameter matrix $G$ is forced to be symmetric and
positive-definite by expressing it as
\beq
	\Gij = \zi \cdot \zj = \sum_{\mu=1}^{N-1} z_{i\mu} z_{j\mu}
\eeq

The general expression for the variational free energy is
\beq
	\hat{F} = -T \log Z_V - \langle E_V \rangle_V + \langle E
	\rangle_V
\eeq
where the expectation values are with respect to the normalized
variational Boltzmann distribution $\exp(-E_V/T)/Z_V$.  The first two
terms are trivial to compute. The first is
\beq
	 -T \log Z_V = \frac{3T}{2} \log \det \Ginv \equiv -3T \log
	\det z
\eeq
apart from a trivial constant that can be neglected, as can the second term,
\beq
	 - \langle E_V \rangle_V = -\frac{3}{2}(N-1)T
\eeq

The last term,
\beq
	\langle E \rangle_V = \half \sum_i \langle \ri^2
	\rangle_V + \sumsig \left \langle \frac{1}{r_{\sig}}
	\right \rangle_V
\eeq
consists in a sum of terms, each amounting to the variational
expectation value of a simple function of a Gaussian vector variable
$\rsig$, of which $\ri$ is a special case. Its probability
distribution is given by
\beq
\label{Px}
	P(\rsig) \propto \exp \left ( -\frac{\rsig^2}{2
	\zsig^2} \right )
\eeq
with
\beq
	\zsig \equiv \sumisig \zi
\eeq
Thus, we have
\beq
\label{z2}
	\langle \ri^2 \rangle_V = 3 \zi^2
\eeq
and
\beq
	\left \langle \frac{1}{r_{\sig}} \right \rangle_V =
	\sqtopi \frac{1}{z_{\sig}} \equiv
	U^C_1(z_{\sig})
\eeq

Summing up, the variational free energy takes the form
\beq
	\hat{F} = -3T \log \det z + \frac{3}{2} \sum_i \zi^2 +
	\sumsig U^C_1(z_{\sig})
\eeq
to be minimized with respect to the variational parameters $\zi$.

\subsubsection*{Gaussian and Translational Parameters}

For the more general variational Ansatz with additional translational
parameters $\ai$,
\beq
\label{EVA}
	E_V/T = \half \sum_{i,j} \Ginvij \, (\ri - \ai) \cdot
	(\rj - \aj)
\eeq
the main difference is a translation of the individual probability
distributions of eq. (\ref{Px}), which now read
\beq
	P(\rsig) \propto \exp \left (
	-\frac{(\rsig-\asig)^2}{2 \zsig^2} \right )
\eeq
with
\beq
	\asig \equiv \sumisig \ai
\eeq
This gives
\beq
\label{z2a2}
	\langle \ri^2 \rangle_V = 3 \zi^2 + \ai^2
\eeq
and
\beq
	\left \langle \frac{1}{r_{\sig}} \right \rangle_V =
	\frac{1}{a_{\sig}} \erf \left ( \frac{a_{\sig}}{\sqrt{2}
	z_{\sig}} \right ) \equiv U^C_2(z_{\sig},a_{\sig})
\eeq
The variational free energy becomes
\beq
	\hat{F} = -3T \log \det z + \half \sum_i (3 \zi^2 +
	\ai^2 ) + \sumsig U^C_2(z_{\sig},a_{\sig})
\eeq

\subsection*{Screened Coulomb Chain}

Next we consider the Debye-screened version of the Coulumb chain, with the
energy
\beq
\label{Escr}
	E = \half \sum_i \ri^2 + \sumsig
	\frac{\exp(-\kappa r_{\sig})}{r_{\sig}}
\eeq

\subsubsection*{Gaussian Parameters Only}

For the variational Ansatz with only Gaussian parameters, eq.
(\ref{PG}), we need eq. (\ref{z2}) and the expectation value
\beq
	\left \langle \frac{\exp(-\kappa r_{\sig})}{r_{\sig}}
	\right \rangle_V = \sqtopi \frac{1}{z_{\sig}} -
	\kappa \exp \left ( \frac{\kappa^2 z_{\sig}^2}{2} \right )
	\erfc \left ( \frac{\kappa z_{\sig}}{\sqrt{2}} \right )
	\equiv U^D_1(z_{\sig})
\eeq
The variational free energy then reads
\beq
	\hat{F} = -3T \log \det z + \frac{3}{2} \sum_i \zi^2 +
	\sumsig U^D_1(z_{\sig})
\eeq

\subsubsection*{Gaussian and Translational Parameters}

Finally, if for the screened Coulomb chain, eq. (\ref{Escr}), also
translational parameters are used in $E_V$, eq. (\ref{EVA}), we will
need the following result in addition to eq. (\ref{z2a2}),
\beqa
	\nonumber \left \langle \frac{\exp(-\kappa
	r_{\sig})}{r_{\sig}} \right \rangle_V & = &
	\frac{\exp(\kappa^2 z_{\sig}^2 / 2)}{2a_{\sig}} \left [
	\exp(-\kappa a_{\sig}) \erfc \left ( \frac{\kappa
	z_{\sig}^2 - a_{\sig}}{\sqrt{2}z_{\sig}} \right ) -
	\exp(\kappa a_{\sig}) \erfc \left ( \frac{\kappa
	z_{\sig}^2 + a_{\sig}}{\sqrt{2}z_{\sig}} \right ) \right
	] \\ & \equiv & U^D_2(z_{\sig},a_{\sig})
\eeqa
The variational free energy will read
\beq
	\hat{F} = -3T \log \det z + \half \sum_i ( 3 \zi^2 +
	\ai^2 ) + \sumsig U^D_2(z_{\sig},a_{\sig})
\eeq

\newpage
\section{The Virial Identity}
\setcounter{equation}{0}

In this appendix we derive the virial identities and show that these
are respected by the variational approach.

\subsection*{Exact Virial Identity}

For any system described by a Boltzmann distribution
\beq
\label{Egen}
	P(\x) = \frac{1}{Z} \exp(-E(\x)/T)
\eeq
with $\x \in {\cal R}^D$, and $E$ rising as a power for large $|\x|$,
we will have
\beq
	\frac{1}{Z} \int \nabla \cdot ({\bf f}(\x) \exp(-E/T)) dx = 0
\eeq
for e.g. any polynomial ${\bf f}$, due to the integrand being an exact
divergence.  This is equivalent to
\beq
\label{f}
	T \langle \nabla \cdot {\bf f} \rangle = \langle {\bf f} \cdot \nabla E
\rangle
\eeq
Thus, by varying ${\bf f}$, we can obtain an infinite set of
identities for the system.

The {\em virial identity} results from the particular choice ${\bf f}
= \x$; in its general form it reads
\beq
	\langle \x \cdot \nabla E \rangle = T D
\eeq
where $D$ is the dimension of $\x$-space ($\x \cdot \nabla$ is the
{\em scaling operator}).

This is particularly useful if the energy $E$ is given by a sum of
terms $E_a$ homogeneous in $\x$,
\beq
	\x \cdot \nabla E_a = \lambda_a E_a
\eeq
in which case the virial identity takes the simple form
\beq
	\sum_a \lambda_a \langle E_a \rangle = T D
\eeq

This applies e.g. to the case of the unscreened polyelectrolyte. There
the scaling operator is given, in relative coordinates, by $\sum_i \ri
\cdot \nabla_{\ri}$, and we have $\lambda_G=2$ and $\lambda_C=-1$; the
virial identity thus reads
\beq
\label{varvir}
	2 \langle E_G\rangle - \langle E_C\rangle = 3(N-1)T
\eeq

\subsection*{Variational Virial Identity}

The virial identity is preserved by the variational approach under
certain conditions, to be specified below.  For the generic system
above, minimizing the free energy
\beq
	\hat{F} = F_V + \langle E - E_V \rangle_V
\eeq
w.r.t. the parameters $\alpha_i$ of a variational energy
$E_V$, leads to (cf. eq. (\ref{delF}) in Appendix A)
\beq
\label{due}
	-T \frac{\partial \hat{F}}{\partial \alpha_i} \equiv \left \langle (E-E_V)
	\frac{\partial E_V}{\partial \alpha_i} \right \rangle^V_C = 0
\eeq
Now, choosing ${\bf f} = \x(E - E_V)$ in eq. (\ref{f}) with the
variational Boltzmann distribution, we have
\beq
\label{DTE}
	DT \langle E-E_V \rangle_V + T \langle \x \cdot \nabla (E -
	E_V) \rangle_V = \langle(E-E_V) \x \cdot \nabla E_V \rangle_V
\eeq
On the other hand, since the virial theorem holds for $E_V$,
\beq
	DT = \langle \x \cdot \nabla E_V \rangle_V
\eeq
Substituting this into eq. (\ref{DTE}), we obtain
\beq
\label{rhs}
	T \langle \x \cdot \nabla (E - E_V) \rangle_V = \langle(E-E_V) \x \cdot
	\nabla E_V \rangle^V_C
\eeq

If the set of parameters $\alpha_i$ of $E_V$ is such (and this is the crucial
condition), that the scaling operation on $E_V$ can be written in terms of
derivatives with respect to the parameters, i.e. if
\beq
	\x \cdot \nabla E_V = \sum_i G_i(\alpha) \frac{\partial E_V}{\partial
	\alpha_i}
\eeq
then the righthand side of eq. (\ref{rhs}) vanishes at the minimum, due to
eq. (\ref{due}), and we are left with
\beq
	\langle \x \cdot \nabla E \rangle_V = \langle \x \cdot \nabla E_V
\rangle_V = DT
\eeq
which is what we desired.

Note that the derivation only relies on a local extremum of the free
energy. Thus, for the polymer, the virial identity,
eq. (\ref{varvir}), is respected by both the rigid ($\a \neq 0$) and
the purely fluctuating ($\a = 0$) solutions.

\newpage
\section{High and Low T Expansions}
\setcounter{equation}{0}

\subsection*{High T expansions}

\subsubsection*{Exact results}

At high $T$, the chain size will be large, and accordingly, the Gaussian term
will dominate over the interaction $V$ in the energy expression,
\beq
	E = E_G + V = \frac{1}{2}\sum_i r_i^2 + \sumsig v(r_{\sig})
\eeq
with $\sig$ summed over contiguous subchains.

It is then natural to attempt an expansion in the perturbation $V$.  For an
arbitrary expectation value, we have the perturbative expansion
\beq
	\langle f \rangle = \langle f \rangle^0 - \frac{1}{T} \langle f V
	\rangle^0_C + \frac{1}{2T^2} \langle f V V \rangle^0_C - \ldots
\eeq
where $\langle \; \rangle^0_C$ refers to connected expectation values
(cumulants) in the unperturbed Boltzmann distribution.  Due to the singular
behaviour of $v(r)$ for small $r$ in the (screened or unscreened) Coulomb case,
only the first few terms will be finite. This indicates that expectation values
cannot be expanded in a pure power series in $T$, and that logarithmic
corrections will occur after the first finite terms.

We are interested in quadratic expectation values of the type $\langle
\ri \cdot \rj\rangle$.  These can be combined to give e.g. the rms
end-to-end distance $r_{ee}$, the gyration radius, and the Gaussian energy
$\langle E_G \rangle$ (and thereby, in the pure Coulomb case, the interaction
energy $\langle E_C \rangle$ by the virial identity).

For the pure Coulomb chain, $v(r) = 1/r$, we get the perturbative expansion
\beq
\label{hiT_C}
	\langle \ri \cdot \rj\rangle = 3 T \delta_{ij} + \sqrt{\frac{2}{\pi T}}
	\sumsigij L_{\sig}^{-3/2} + O(T^{-2})
\eeq
where $\sig$ denotes a contiguous sub-chain containing the $i$th and the $j$th
bond, and $L_{\sig}$ its total number of bonds.  This leads to
\beq
	\langle E_G \rangle = 3(N-1)T/2 + \sqrt{\frac{1}{2 \pi T}} \sumsig
	L_{\sig}^{-1/2} + O(T^{-2})
\eeq
and
\beq
	r_{ee}^2 = 3(N-1)T + \sqrt{\frac{2}{\pi T}} \sumsig L_{\sig}^{1/2} +
	O(T^{-2})
\eeq

Similar results are obtained for the case of a screened Coulomb interaction,
$v(r) = e^{-Kr}/r$, where the expansion of a quadratic expectation value gives
\beq
\label{hiT_D}
	\langle \ri \cdot \rj\rangle = 3 T \delta_{ij} + \frac{3}{\kappa^2T}
	\sqrt{\frac{2}{\pi T}} \sumsigij L_{\sig}^{-5/2} + O(T^{-3})
\eeq

\subsubsection*{Variational results}

The corresponding variational results can also be expanded at high $T$ (where
$\ai=0$), by expanding the variational solution $z_{i\mu}$ around the
unperturbed value, which can be chosen as $\sqrt{T} \delta_{i\mu}$.  The
variational approximation to a quadratic expectation value, $\langle \ri \cdot
\rj \rangle_V = 3 \zi \cdot \zj$, can then easily be expanded. The results thus
obtained reproduce the exact results, eqs. (\ref{hiT_C},\ref{hiT_D}), correctly
to the order shown.

We conclude that, independently of screening, the high $T$ variational results
are correct to next-to-leading order for $E$ and $E_G$, and to leading order for
$E_C$.

\subsection*{Low T expansions}

Here we will treat only the pure Coulomb case in detail; most of the discussion
applies also to the screened case.

\subsubsection*{Exact results}

At low $T$, expectation values can be expanded around the configuration that
minimizes the energy.  This expansion is slightly complicated by the the
rotational degeneracy of the minimum.  The results can be expressed in terms of
the classical configuration, which is given by a straight line configuration.

Let $b_i$ be the the bond-lengths at the energy minimum. These have to be
computed numerically, by solving the equation
\beq
\label{b_i}
	b_i = \sumsigi \frac{1}{b_{\sig}^2}
\eeq
where $b_{\sig}$ is the length of a subchain containing the $i$th bond.  Note
that the above equation can be written as a matrix equation:
\beq
	b_i = \sum_j b_j \sumsigij \frac{1}{b_{\sig}^3} = \sum_j B_{ij} b_j
\eeq
Thus, $b$ is an eigenvector of the matrix $B$ with a unit eigenvalue. Similarly,
we can define a whole series of tensors:
\beqa
	E_C & = & \sumsig \frac{1}{b_{\sig}} \\ A_i & = & \sumsigi
	\frac{1}{b_{\sig}^2} \equiv b_i \\ B_{ij} & = & \sumsigij
	\frac{1}{b_{\sig}^3} \\ C_{ijk} & = & \sumsigijk \frac{1}{b_{\sig}^4}
\eeqa
They are all symmetric, and contracting either with $b_i$ gives the tensor of
rank one less.

In addition, we need two more matrices, related to $B$,
\beqa
	U & = & (1 + 2B)^{-1} \\ V & = & P(1 - B)^{-1}P
\eeqa
where $P$ denotes the projection matrix onto the subspace orthogonal to $b$,
which is deleted by $1-B$.

In terms of these tensors, we have the quadratic expectation-values at low $T$:
\beq
	\langle \ri \cdot \rj \rangle = b_i b_j + T \left ( U_{ij} + 2 V_{ij} +
	\frac{4 b_i b_j}{3\sum_k b_k^2} + 3 \sum_{klm} C_{klm} (b_i U_{jk} + b_j
	U_{ik}) ( U_{lm} - V_{lm} ) \right ) + O(T^2)
\eeq
where the first two terms of the $T$ coefficient are the naive contributions
from the longitudinal and transverse fluctuations.  The rest are corrections due
to the rotational degeneracy of the $T=0$ configuration, which is also
responsible for the transverse zero-modes (of $1-B$).

From this we obtain e.g the average Gaussian energy,
\beq
	\langle E_G \rangle = 1/3 E_0 + T(3 N / 2 - 11 / 6) + O(T^2)q
\eeq
from which, using the virial identity, we obtain
\beq
	\langle E_C \rangle = 2/3 E_0 - 2 T/ 3 + O(T^2)
\eeq
and
\beq
\label{EPG}
	\langle E \rangle = E_0 + T (3 N - 5) / 2 +
	O(T^2)
\eeq
where $E_0 = 3/2 \sum_i b_i^2$ is the exact energy at $T=0$.  In the
last equation, the $T$-coefficient is, as it should, half the number
of degrees of freedom, not counting the two rotational zero-modes (and
the three translational ones already removed).

In the screened case, similar results can be obtained. In particular,
eq. (\ref{EPG}) remains valid, though with a different $E_0$.

\subsubsection*{Variational results}

Similarly, the variational results can be expanded at low $T$. We have
to distinguish between the two different solutions.

For the {\em purely fluctuating} solution with $\ai = 0$, the
variational free energy is, at $T=0$,
\beq
	\hat{F}_0 \equiv \langle E \rangle_V = \frac{3}{2} \sum_i
	\zi^2 + \sqtopi \sumsig \frac{1}{z_{\sig}}
\eeq
This is obviously just the energy of an $(N-1)$-dimensional version of
the chain, with modified coefficients.  It is minimized by the aligned
configuration
\beq
	\zi = \left ( \frac{2}{9 \pi} \right )^{1/6} b_i \hat{\bf n}
\eeq
where $b_i$ are given by eq. (\ref{b_i}), and $\hat{\bf n}$ is some
unit vector in $N-1$ dimensions.  For small but finite $T$, we have to
add the entropy term, $-3T\log \mbox{det}
\{\z\}$, to $\hat{F}_0$.  This forces the configuration out of alignment. The
resulting configuration can be obtained as a low-$T$ expansion around
the $T=0$ solution. The first correction to $\z$ will be of order
$\sqrt{T}$, and since the $T=0$ $\zi $ are aligned, the matrix inverse
$\wi$ diverges - it will go like $1/\sqrt{T}$.

For the total energy, the leading correction can be obtained as
follows. The equation to solve is
\beq
	3T\wi = \nabla_i \hat{F}_0(\z)
\eeq
For the $T=0$ solution, $\nabla_i \hat{F}_0(\z_0) = 0$. Thus, to lowest order,
\beq
	3T\wi = \nabla_i \sum_j \nabla_j \hat{F}_0(\z_0) \cdot d\zj
\eeq
The leading energy correction will be
\beq
	d\hat{F}_0 = \half \sum_{ij} \nabla_i \nabla_j
	\hat{F}_0(\z_0) d\zj d\zi = \frac{3T}{2} \sum_i \wi \cdot d\zi
\eeq
Now, $d\zi$ consists of an aligned part and a transverse part, both
$\propto \sqrt{T}$, while for $\wi$ the aligned part is $\propto 1$,
while the transverse part is $\propto 1/\sqrt{T}$.  Thus, the leading
contribution to $d\hat{F}_0$ comes from the transverse part. Taking
the trace of the tranverse part of the identity $\wi \cdot \zj =
\delta_{ij}$, and noting that the transverse part of $\z$ sits
entirely in $d\z$, we have to leading order
\beq
\label{DF0}
	d\hat{F}_0 = \frac{3T(N - 2)}{2} \equiv d\langle E \rangle_V
\eeq

Using the virial identity, which holds also for the variational
expectation values, we get to first order in $T$:
\beqa
	\langle E \rangle_V & = & (\frac{6}{\pi})^{1/3} E_0 +
	\frac{3T(N - 2)}{2} + O(T^2)
\\
	\langle E_G \rangle_V & = & \frac{1}{3} (\frac{6}{\pi})^{1/3}
	E_0 + \frac{T(3N - 4)}{2} + O(T^2)
\\
	\langle E_C \rangle_V & = & \frac{2}{3} (\frac{6}{\pi})^{1/3}
	E_0 - T + O(T^2)
\eeqa
where $E_0$ is the exact $T=0$ energy.  Note that already the
zero-order results are off by a factor $(\frac{6}{\pi})^{1/3}
\approx 1.24$, but that the correction to $E$ is correct in the high
$N$ limit.

Similar results are obtained for the screened Coulomb chain. Most of
the general analysis leading to eq.(\ref{DF0}) still holds, and we
have e.g.
\beq
	\langle E \rangle_V = E_0' + \frac{3T(N - 2)}{2} + O(T^2)
\eeq
with $E_0' \neq E_0$.

For the {\em symmetry-broken} $\ai \neq 0$ solution, the $T=0$
configuration is instead given by
\beq
	\ai = b_i \hat{\bf n} \; , \; \zi = 0
\eeq
with $b_i$ as in eq. (\ref{b_i}), and $\hat{\bf n}$ an arbitrary unit
vector in ${\cal R}^3$.  The small $T$ corrections are obtained from
expressing $\hat{F}$ as
\beq
	\hat{F} = -3T\log \mbox{det} \{\z\} + \langle E(\ai +
	\sum_{\mu} z_{i\mu} \J_{\mu}) \rangle_J
\eeq
where $\J_{\mu}$ are uncorrelated standard Gaussian noise variables. Expanding
in $\z$, we obtain
\beq
	\hat{F} = -3T\log \mbox{det} \{\z\} + E(\a) + \half \sum_{ij}
	\zi \cdot \zj \nabla_i \cdot \nabla_j E(\a) + \ldots
\eeq
Because the Coulomb term satisfies Laplace' equation, this is just
(provided $\a \neq 0$)
\beq
	\hat{F} = -3T\log \mbox{det} \{\z\} + E(\a) + \frac{3}{2}
	\sum_i \zi \zi
\eeq
and the variational free energy separates in $\z$ and $\a$ for small
$\z$.  Thus, minimum is obtained for
\beq
	\ai = \hat{\bf n} b_i
\eeq
and $\z$ satisfies
\beq
	-3T \wi + 3 \zi = 0
\eeq
which means
\beq
	\zi \cdot \zj = T \delta_{ij}
\eeq
The energies become
\beqa
	\langle E \rangle_V & = & E_0 + \frac{3T(N - 1)}{2} \\
	\langle E_G \rangle_V & = & \frac{1}{3} E_0 + \frac{3T(N - 1)}{2} \\
	\langle E_C \rangle_V & = & \frac{2}{3} E_0
\eeqa
which is correct to lowest order, and for large $N$ qualitatively
correct to first order (except for $E_C$).

Again, the screened chain lead to similar results; in particular, the
$T=0$ energies will be the correct ones.

\newpage
\section{Zero Temperature Scaling Properties}
\setcounter{equation}{0}

The $T=0$ configuration of a pure Coulombic chain cannot be obtained
analytically, but must be computed numerically. However, an approximate
calculation can be done. The equation for the bond lengths $b_i$ in the
elongated ground state configuration is given by eq. (\ref{b_i}):
\beq
	b_i = \sumsigi \frac{1}{b_{\sig}^2}
\eeq
By assuming that the bond length is locally approximately constant,
this can be approximated by
\beq
	b_i \approx \frac{1}{b_i^2} \sum_{k=1}^i \sum_{l=i}^{N-1} (l-k+1)^{-2}
\eeq
This can be rewritten as
\beq
	b_i^3 \approx \sum_{l=1}^{N-1}(-l+\mbox{min}(l,N-i)+\mbox{min}(l,i))/l^2
\eeq
This in turn can be approximated by an integral, leading to
\beq
\label{b_i1}
	b_i \approx \left [ \log \left ( \mbox{const}\frac{i(N-i)}{N}
	\right ) \right ]^{1/3}
\eeq
Defining $s = i/N$, this amounts to
\beq
\label{b_i2}
	b_i \approx \left ( \log(\mbox{const} \; N s (1 - s) ) \right )^{1/3}
\eeq

Eqs. (\ref{b_i1},\ref{b_i2}) give a quite accurate picture of the variation of
the bond lengths at $T=0$. They also imply, that the typical ground state
bond length should swell roughly as $(\log N)^{1/3}$ for large $N$.

\end{document}